\documentclass[lettersize,journal,nosections]{IEEEtran}
\usepackage{amsmath,amsfonts}
\usepackage{mathabx}
\usepackage[linesnumbered,ruled,vlined]{algorithm2e}
\usepackage{array}
\pdfoutput=1
\usepackage[caption=false,font=normalsize,labelfont=sf,textfont=sf]{subfig}
\usepackage{textcomp}
\usepackage{stfloats}
\usepackage{url}
\usepackage{verbatim}
\usepackage{graphicx}
\usepackage{cite}
\usepackage{nameref}
\usepackage{enumitem}
\usepackage{booktabs}
\usepackage{cleveref}
\crefformat{table}{#2Table~\Roman{#1}#3}
\usepackage[inkscapelatex=false]{svg}
\usepackage{threeparttable}
\usepackage{multirow}
\usepackage[ruled,vlined]{algorithm2e}
\SetKwFor{ParFor}{parallel for}{do}{end} 
\usepackage{fancyhdr} 
\usepackage[switch]{lineno}
\usepackage{xcolor} 
\usepackage{subcaption} 

\usepackage{graphicx}
\usepackage{subfig} 
\usepackage{tikz}
\newcommand*\circled[1]{\tikz[baseline=(char.base)]{
            \node[shape=circle,draw,fill=black,text=white,inner sep=1pt] (char) {\footnotesize #1};}}

\begin{document}
\title{Critical Path Aware Timing-Driven Global Placement for Large-Scale Heterogeneous FPGAs}

\author{
    He Jiang$^{\ast}$, Yi Guo, Shikai Guo$^{\ast}$, Huijiang Liu, Xiaochen Li, Ning Wang, Zhixiong Di

    \IEEEcompsocitemizethanks{
    
    \IEEEcompsocthanksitem H Jiang, Y Guo, H. Liu, and X. Li are with the School of Software, Dalian University of Technology, Dalian, China. E-mail: jianghe@dlut.edu.cn, gy2220212088@dlmu.edu.cn,
    liuhuijiang@mail.dlut.edu.cn, xiaochen.li@dlut.edu.cn.

    \IEEEcompsocthanksitem S. Guo is with the School of Information Science and Technology, Dalian Maritime University, Dalian, China. E-mail: shikai.guo@dlmu.edu.cn.

    \IEEEcompsocthanksitem N Wang is with Chengdu Sino Microelectronics System Co.Ltd., Chengdu 610031, China. E-mail: wangning@csmsc.com.

    \IEEEcompsocthanksitem Z. Di is with the School of Information Science and Technology, Southwest Jiaotong University, Chengdu 610031, China. E-mail: zxdi@home.swjtu.edu.cn.

    \IEEEcompsocthanksitem * Corresponding author (He Jiang and Shikai Guo)
    
    }
	
}


\markboth{IEEE Transactions on Computer-Aided Design of Integrated Circuits and Systems,~Vol.~1, No.~1, July~2025}%
{Shell \MakeLowercase{\textit{Guo et al.}}: A Sample Article Using IEEEtran.cls for IEEE Journals}


\maketitle
\begin{abstract}
Timing optimization during global placement is critical for achieving optimal circuit performance and remains a key challenge in modern Field Programmable Gate Array (FPGA) design. 
As FPGA designs scale and heterogeneous resources increase, dense interconnects introduce significant resistive and capacitive effects, making timing closure increasingly difficult.
Existing methods face challenges in constructing accurate timing models due to multi-factor nonlinear constraints as well as load and crosstalk coupling effects arising in multi-pin driving scenarios.
To address these challenges, we propose TD-Placer, a critical path aware, timing-driven global placement framework. It leverages graph-based representations to capture global net interactions and employs a nonlinear model to integrate diverse timing-related features for precise delay prediction, thereby improving the overall placement quality for FPGAs.
TD-Placer adopts a quadratic placement objective that minimizes wirelength while incorporating a timing term constructed by a lightweight algorithm, enabling efficient and high-quality timing optimization.
Regarding net-level timing contention, it also employs a finer-grained weighting scheme to facilitate smooth reduction of the Critical Path Delay (CPD).
Extensive experiments were carried out on seven real-world open-source FPGA projects with LUT counts ranging from 60K to 400K.
The results demonstrate that TD-Placer achieves an average $\sim$10\% improvement in Worst Negative Slack (WNS) and a $\sim$5\% reduction in CPD compared to the state-of-the-art method, with an average CPD comparable ($\times 1.01$) to the commercial AMD Vivado across five versions (2020.2–2024.2). Its code and dataset are publicly available\footnote{\url{https://github.com/yyiloe/TD-Placer}}.
\end{abstract}


\section{\textbf{Introduction}}
\label{intro}

Field Programmable Gate Array (FPGA) is a programmable platform for flexible user customization, featuring short development cycles and zero non-recurring engineering costs \cite{lin2021timing}.
As user demands diversify, FPGAs grow in capacity and heterogeneity to support more complex designs, placing greater demands on placement tools in both runtime and solution quality \cite{elgamma2020learn}.
Specifically, placement, which maps logic blocks to physical locations on the FPGA, directly determines routability in the subsequent routing stage and significantly affects circuit wirelength and timing quality\cite{liang2024amf, xiong2024data, chang2000timing}.

To deliver high-quality placement solutions across diverse application scenarios, two placement methods are widely adopted in academia and industry: simulated annealing \cite{murray2020vtr} and analytical placement\cite{gort2012analytical}. 
Simulated annealing iteratively swaps block positions to optimize wirelength under legalization constraints. When targeting larger netlists, analytical placement methods provide a better trade-off between placement quality and runtime, and demonstrate stronger scalability \cite{mai2023openparf}. A typical analytical placement flow consists of three phases: global placement, legalization, and detailed placement \cite{dhar2017effective}. Global placement, which roughly positions elements under resource constraints, plays a crucial role in shaping instance distribution and greatly impacts final placement quality.

Traditional analytical placers minimize total wirelength while ensuring routability of the solution in the global placement stage, indirectly addressing timing constraints by approximating net delay using wirelength metrics \cite{gengjie2018ripplefpga,pattison2016gplace,kuo2017clock,li2017utplacef}. 
However, as modern FPGA designs continue to scale in complexity and heterogeneous resource integration, the relationship between FPGA interconnect delays and wirelength has become highly nonlinear\cite{lin2021timing}. 
Consequently, Wirelength-driven techniques increasingly fail to achieve optimal timing quality, or even ensure timing closure \cite{chang2000timing}. 
To address this challenge, timing-driven global placement has been extensively studied. It integrates the timing model into the global placement loop, dynamically estimates post-routing timing, and models the interplay between wirelength, routability, and timing to improve circuit performance. Lin et al. \cite{lin2021timing} developed an timing model based on FPGA routing architecture and net delay mapping, incorporating clock and density costs during global placement to ensure clock resource legality.
Liang et al. \cite{liang2024amf} proposed a hard-coded piecewise polynomial model for critical path identification, and applied a series of pseudo-net strategies during placement iterations to optimize the timing.
Xiong et al. \cite{xiong2024data} proposed a lightweight data-driven linear timing model with routing congestion estimates, integrated into an electrostatic global placement scheme to optimize timing.

However, as design scales increase and the complexity of heterogeneous resources and architectures in modern FPGAs persists, the resistive and capacitive effects introduced by
dense heterogeneous interconnects make net delays more sensitive to a wider range of timing-related features and the global context of the net. 
Therefore, to enhance the quality of timing-driven placement, it is essential to address two fundamental challenges in current methods.

\textbf{Challenge 1: Lacking awareness of load and crosstalk coupling effects in multi-pin driving scenarios.} 
Each FPGA interconnect net consists of one driver pin and multiple load pins, which share a portion of the routing path.
The delay between any driver-load pin pair can vary significantly due to load and crosstalk coupling effects from other instances on the net, which limits the accuracy of existing single-pin pair models.
Therefore, a key challenge is how to incorporate the full net topology for accurate timing analysis.

\textbf{Challenge 2: Lacking nonlinear modeling of diverse timing features affecting net delays.} 
In modern FPGA architectures, net delays arise from nonlinear interactions among factors such as programmable switches, segmented routing, heterogeneous instance cascades, geometric distance, parasitic resistance, capacitance, and inductance. 
These factors make accurate net delay prediction difficult, and inaccurate estimates can misguide placement, significantly degrading post-routing timing quality. 
Therefore, developing a model that comprehensively captures these nonlinear timing factors is one of the core challenges in timing-driven placement optimization.

To address these challenges, we propose \textbf{TD-Placer}, a critical path aware, timing-driven global placement framework to effectively optimize post-routing circuit timing.
TD-Placer consists of three main components: (1) the net information extraction component, (2) the net delay prediction component and (3) the global placement component. 
First, the net information extraction component addresses crosstalk and coupling effects under multi-pin driving scenarios by employing a specially designed graph-based method to capture net topology, providing critical support for accurate delay prediction. 
Then, the net delay prediction component leverages this topology, integrates various timing-related features, and adopts a nonlinear model to estimate precise end-to-end net delays. 
Finally, the global placement component introduces an end-to-end timing metric via fast slack estimation, which is integrated with a wirelength term into a unified quadratic objective to jointly optimize timing and wirelength. 
Within the global placement component, a fine-grained weighting scheme, guided by logic depth and global timing thresholds, is specifically applied along critical timing paths to enable smooth delay reduction during iterative optimization. 
In addition, TD-Placer employs a lightweight net delay prediction algorithm to ensure efficient deep learning model inference.

We conduct extensive experiments on seven open-source FPGA designs with LUT counts ranging from 60K to 400K (as listed in Table \ref{dataset2}). 
The state-of-the-art methods we compare against include: (1) AMF-placer \cite{liang2024amf}, the state-of-the-art open-source timing-driven placer supporting mixed-size macros for Ultrascale devices, (2) five versions (2020.2-2024.2) of the commercial AMD Xilinx Vivado \cite{Xilinx2022}. 
Compared to AMF-Placer, TD-Placer reduces average Critical Path Delay (CPD) by 4\% and 5\% and improves Worst Negative Slack (WNS) by 7\% and 12\% on designs routed with Vivado 2020.2 and 2021.2, respectively. Compared to Vivado, it achieves comparable average CPD ($\times 1.01$) across five versions.

The main contributions of this paper are as follows:
\begin{itemize}
\item  We propose an accurate timing model for global placement, which comprehensively modeling multi-pin load, crosstalk coupling effects, and multiple timing-related factors to enable precise net delay prediction.

\item  We design a lightweight algorithm to ensure efficient timing analysis during global placement. In addition, we introduce a fine grained timing weighting scheme with awareness of global timing thresholds and logic depth, enabling smooth optimization along critical paths.

\item We conduct extensive experiments to validate TD-Placer. Compared to the state-of-the-art method, it achieves a $\sim$5\% reduction in CPD, a $\sim$10\% improvement in WNS, and a 23\% increase in timing prediction accuracy. We also open-source the code and dataset, including scripts for feature extraction and dataset construction, which can be easily adapted to other devices\footnote{https://github.com/yyiloe/TD-Placer}.
\end{itemize}

\begin{figure}[!t] 
\centering
 \includegraphics[width=0.8\linewidth]{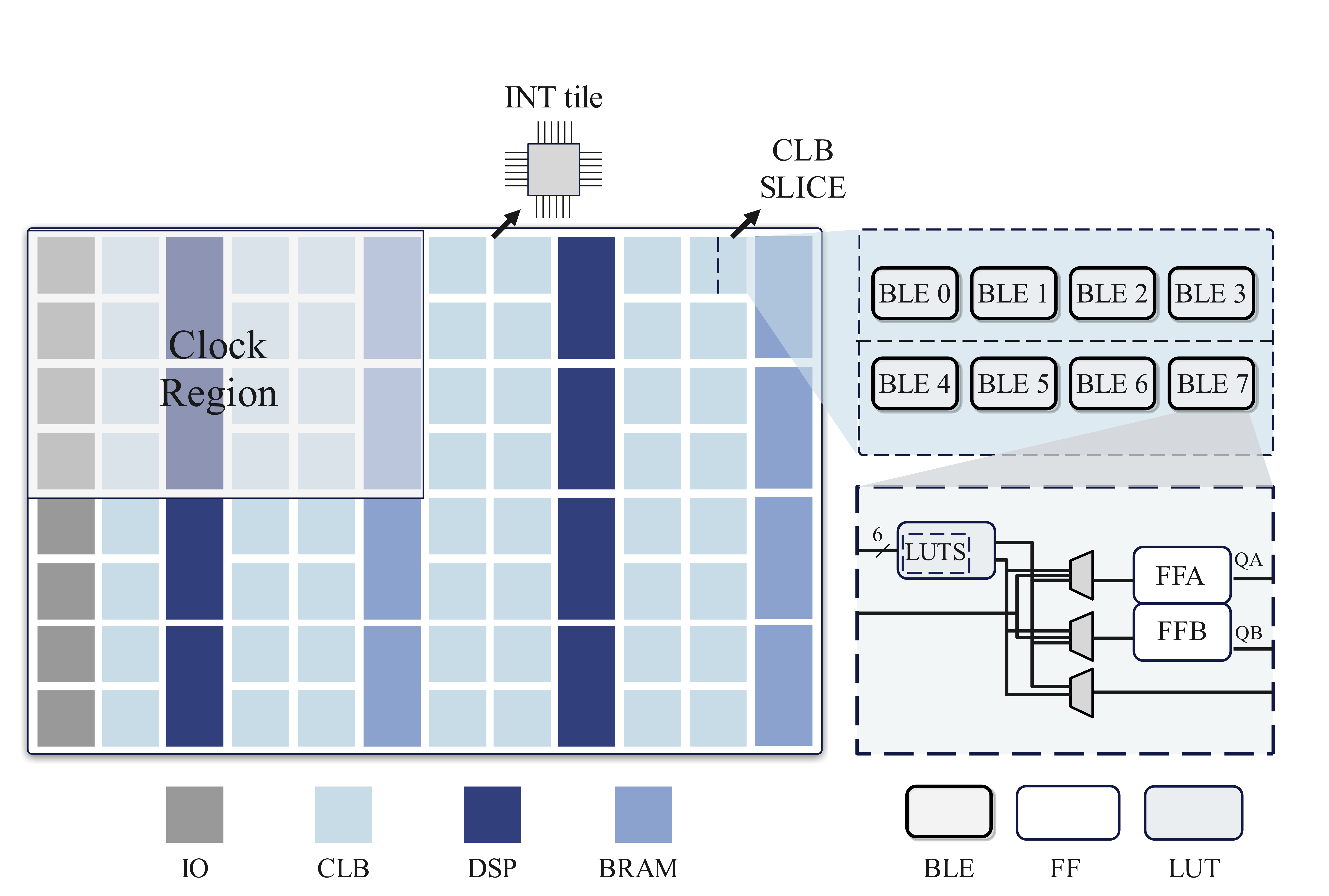}
\caption{Example of a simplified FPGA architecture depiction for Xilinx UltraScale.}
\label{device}
\end{figure}

\section{\textbf{Background and Motivation}}


\subsection{\textbf{FPGA Architecture}}\label{Diff}
We adopt the Xilinx UltraScale architecture VU095 \cite{Xilinx2018} as our target FPGA device. As illustrated in Fig \ref{device}, the architecture consists of a 2D array of configurable sites. Each Configurable Logic Block (CLB) contains two slices, each with 8 Basic Logic Elements (BLEs), and each BLE comprises two Look-Up Tables (LUTs) and two Flip-Flops (FFs). The device also incorporates heterogeneous resources including Digital Signal Processors (DSPs), Block RAMs (BRAMs), and Input/Output (IO) blocks. It comprises a 5×8 array of clock regions, with Fig \ref{device} showing a partial representation. Each clock region accommodates up to 24 distinct clock nets.

FPGA routing typically comprises both intra-site routing and inter-site routing. Inter-site routing is achieved through the nearest interconnect (INT) tiles (represented by the white area with black arrows), where the built-in switch box enables switching between wiring segments in all four directions (up, down, left, right) as well as wire branching for multiple fanouts, connecting to different instances respectively. Intra-site routing handles communication between instances within the same site. The routing delay between any two different sites generally consists of both the intra-site delay and the delay through segmented wires in the INT tiles.


\begin{figure*}[htbp]
    \centering
    \subfloat[]{\includegraphics[height=0.45\columnwidth, keepaspectratio]{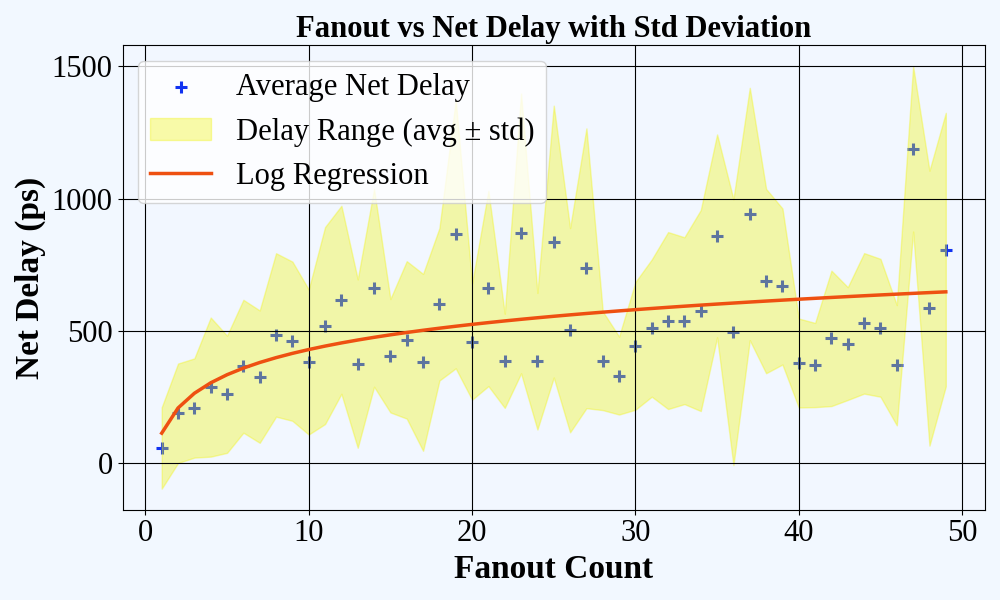}\label{motivation2}}
    \hspace{0.02\linewidth} 
    \subfloat[]{\includegraphics[height=0.45\columnwidth, keepaspectratio]{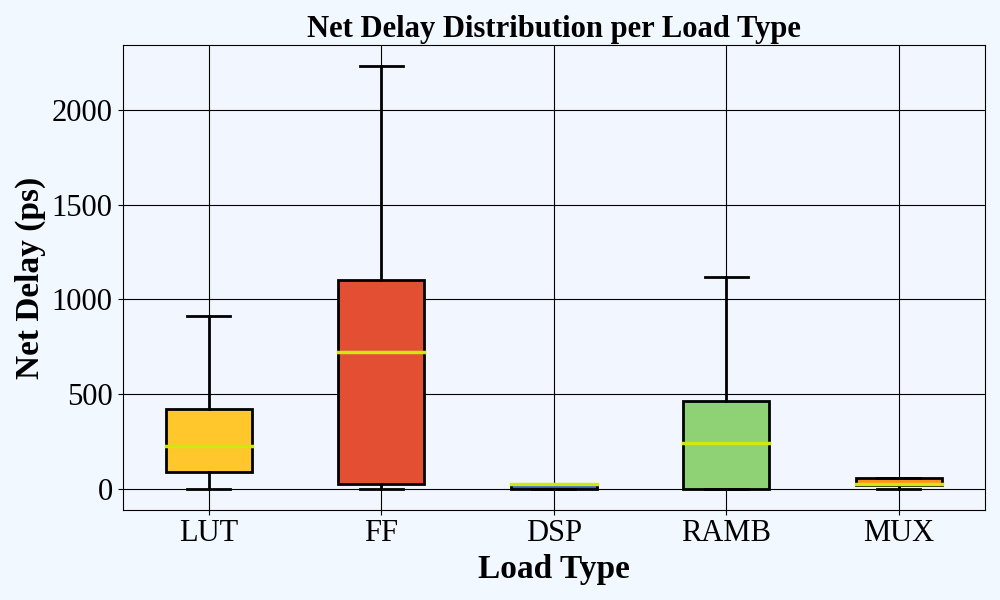}\label{motivation3}}
    \caption{(a) shows a scatter plot illustrating the relationship between net delay and \textbf{net fanout}. (b) presents a box plot of net delay categorized by \textbf{load type}. }
    \label{motivation}
\end{figure*}

\begin{figure}[!t] 
\centering
 \includegraphics[width=0.7\linewidth]{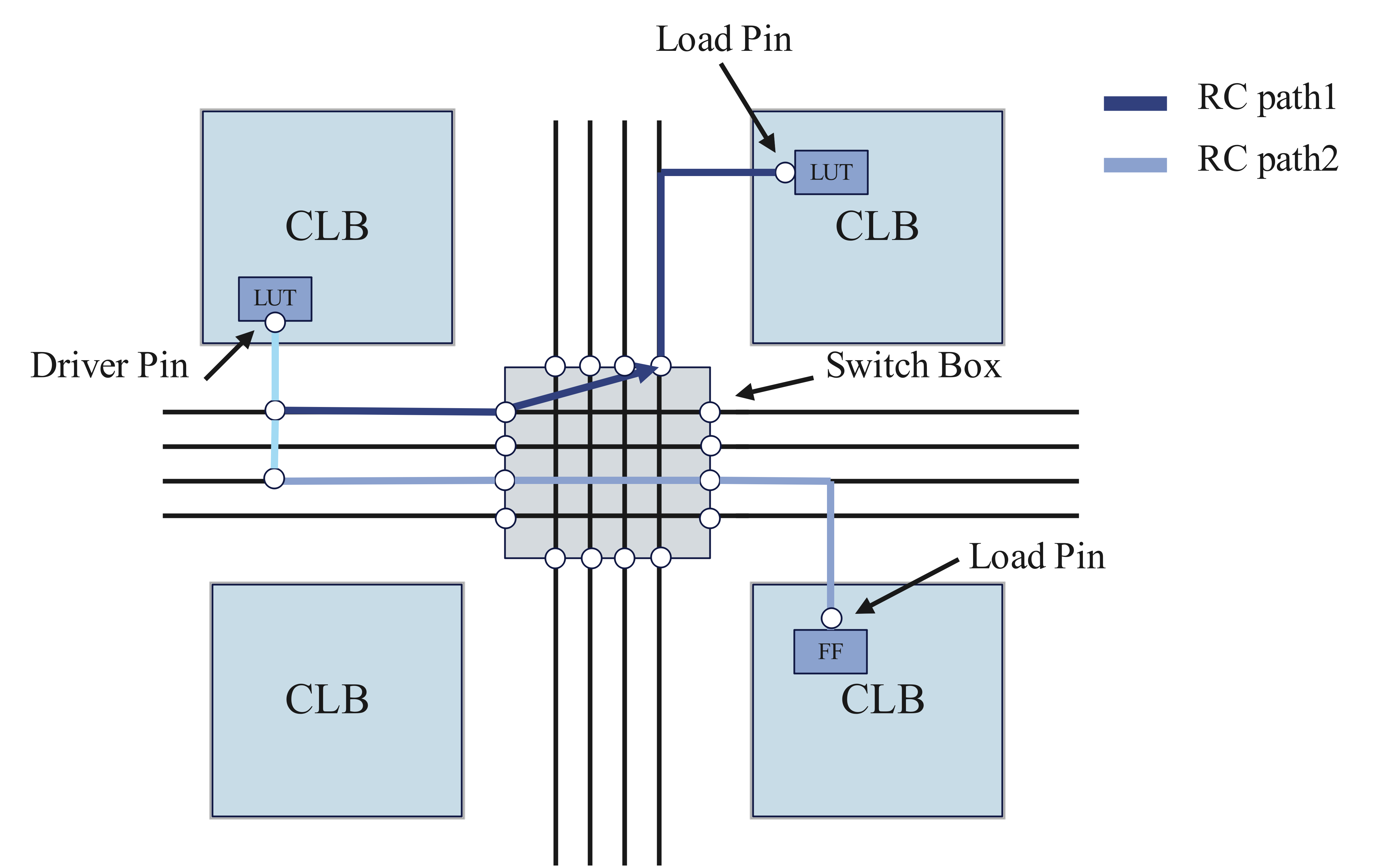}
\caption{The pins in the FPGA drive the RC path of two other pins through the switch box}
\label{motivation1}
\end{figure}








\subsection{\textbf{Traditional Global Placement}}\label{sec:TGP}

The wirelength-driven quadratic placement process determines approximate locations for all cell instances under resource constraints \cite{liang2021amf,pattison2016gplace,wuxi2019new}. To approximate the non-differentiable HPWL (Half-Perimeter Wirelength) function, the Bound2Bound \cite{spindler2008kraftwerk2} net model is employed to weight the quadratic objective function:
\begin{equation}
{W_m} = \sum\limits_{i,j \in m} {\left[ {w_{x,ij}^{B2B}{{\left( {{x_i} - {x_j}} \right)}^2} + w_{y,ij}^{B2B}{{\left( {{y_i} - {y_j}} \right)}^2}} \right]} 
\end{equation}
where $m$ represents a net, $i,j$ denote the  instances connected by the net, \(w_{x,ij}^{B2B}\) and \(w_{y,ij}^{B2B}\) are the weighting coefficients assigned according to the Bound2Bound net model.

To transform the constrained problem into an unconstrained optimization problem, virtual anchors are added to restrict illegal movement of instances \cite{li2017utplacef,gengjie2018ripplefpga,liang2021amf}. Virtual anchors are connected to their associated instances through two-pin pseudo nets:
\begin{equation}
    \mathop {\min }\limits_{x,y} \sum\limits_{m \in Q} {{W_m} + } \sum\limits_{{m_{\rm{p}}} \in {Q_p}} {{w_{{m_p}}}{W_{{m_p}}}(x,y)}
\end{equation}
where ${W_m}$ and ${W_{{m_p}}}$ represent the HPWL approximations for regular nets and pseudo nets respectively, $Q$ and ${Q_p}$ denote the sets of regular nets and pseudo nets (for legalization and density control), ${w_{{m_p}}}$ indicates the weighting coefficients applied to pseudo nets.




\subsection{\textbf{Motivation}}
Net delay is a major contributor to path delay and plays a critical role in circuit timing performance. 
Accurate prediction during placement is essential, as timing analysis relies on repeated delay estimation to guide instance adjustments. 
Since net delay is influenced by multiple factors, predicting delay solely based on geometric distance or simple congestion metrics is insufficient.
As illustrated in Figure \ref{motivation1}, a driver pin in an FPGA net fans out via a switch box to multiple load pins. 
Each programmable switch in the switch box introduces resistance, forming parallel Resistive-Capacitive paths. 
This increases delay, especially with added input capacitance from load types. We analyzed 160,162 nets with driver-load distances shorter than 4\% of the longer dimension of the target layout, as shown in Figure \ref{motivation2}, that average net delay increases significantly from $\sim$200  ps to a peak of $\sim$1200 ps as fanout grows from 1 to 50.
Figure \ref{motivation3} further shows that load types also cause noticeable variations in delay. 
These results highlight that net delay is sensitive to crosstalk, coupling effects, and various timing features, making accurate delay prediction difficult and significantly affecting timing convergence in placement.

\subsection{\textbf{Related Works}}\label{related}
Simulated Annealing (SA), exemplified by the widely used VPR placer \cite{murray2020vtr,rose2012vtr}, effectively handles architectural constraints and non-smooth objectives in FPGA placement, but its convergence efficiency remains limited for large solution spaces despite numerous enhancements \cite{elgamma2020learn,mahmoudi2023respect,murray2019adaptive,elgammal2021rlplace}.

To balance placement quality and runtime, wirelength-driven analytical placers have been widely studied. Chen et al. proposed a bin-packing-based algorithm for large-scale heterogeneous FPGAs, achieving nearly $200\times$ bin-packing acceleration, $3\times$ placement speedup, and 50\% wirelength reduction over VPR 7.0 \cite{chen2014efficient}. Analytical placers from ISPD 2016/2017 contests showed strong results in mitigating routing congestion and clock resource overflow \cite{chen2005ntuplace,li2017utplacef,pattison2016gplace}. Electrostatic placement, modeling device density as charged particles to control congestion, is another key method \cite{li2019elfplace,rajarathnam2022dreamplacefpga}, while OpenParf provides an open-source, GPU-accelerated framework for full physical design flows \cite{mai2023openparf,mai2022multi,wang2023robust}.

With the increasing scale of modern FPGA designs, wirelength-driven placement is insufficient to ensure optimal timing and convergence \cite{chang2000timing}. Timing-driven placers address this by optimizing WNS and TNS through integrated timing models and metrics \cite{lin2021timing, xiong2024data, mai2023multielectrostatic}. Yet, as applications grow more complex, delivering precise timing guidance for densely interconnected devices remains critical and challenging, requiring more advanced algorithms for large-scale heterogeneous circuits.

\begin{figure*}[!t] 
\centering
 \includegraphics[width=0.83\linewidth]{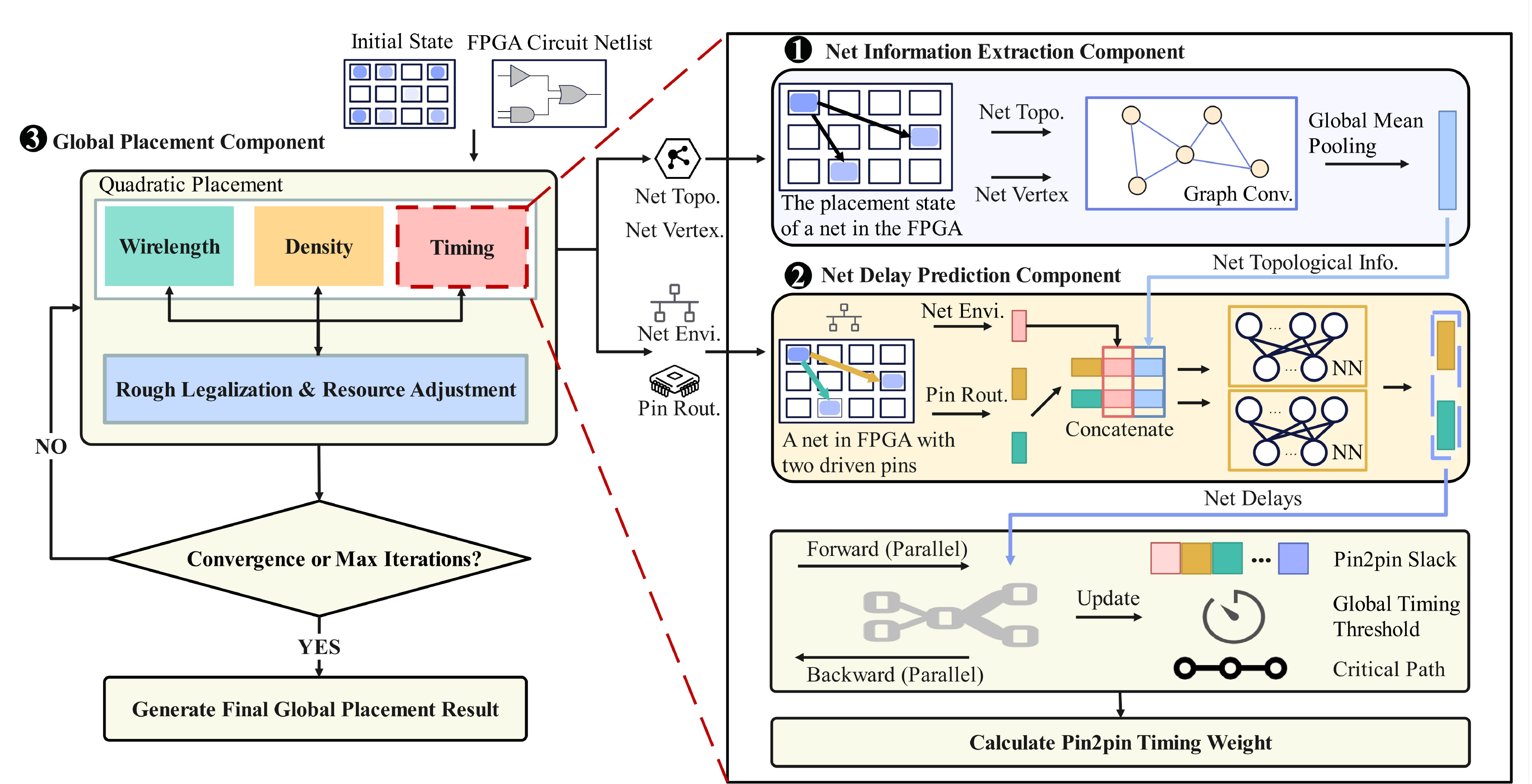}
\caption{The TD-Placer framework.}
\label{framework}
\end{figure*}

\section{\textbf{TD-Placer Framework}}\label{Model}
\subsection{\textbf{Overview}}\label{Overview}

To address the above challenges, we present TD-Placer, a critical path aware, timing-driven global placement framework as illustrated in Figure \ref{framework}. 
TD-Placer comprises three components: \circled{1} the net information extraction component, \circled{2} the net delay prediction component and \circled{3} the global placement component, which collaboratively enable efficient timing optimization and placement convergence.
Firstly, the net information extraction component performs holistic net modeling to provide topology information that capture load and crosstalk coupling effects, enabling more accurate delay prediction.
Then, the extracted net topology information is fed into the net delay prediction component, which integrates other multi-dimensional timing-related features to achieve accurate end-to-end net delay prediction.
Within the global placement component, the hierarchical netlist is converted into a directed acyclic timing graph, with vertex weights set via a logic delay lookup table.
The design’s nets are analyzed to extract topological information for net delay prediction, updating timing arc weights accordingly to prepare for timing analysis.
Hierarchical parallel forward and backward traversals of the timing graph are performed to compute arc-level slacks. 
These slacks, together with the identified critical path and a threshold estimation of global slack, enable the construction of arc-based fine-grained timing metrics to ensure as smooth timing optimization as possible.
Next, these metrics are integrated with wirelength and density terms to improve Worst Negative Slack (WNS), Critical Path Delay (CPD) and Total Negative Slack (TNS). Finally, TD-Placer employs rough legalization and resource adjustment algorithms to enhance placement quality and support subsequent global placement iterations.

\begin{table*}[!t]
  \centering
  \caption{\textsc{Timing-Related Features Used in TD-Placer}}
  \begin{threeparttable}
    \renewcommand{\arraystretch}{1.1} 
    \begin{tabular}{
      >{\centering\arraybackslash}m{1.8cm}   
      >{\centering\arraybackslash}m{0.6cm}   
      >{\centering\arraybackslash}m{2.5cm}   
      >{\arraybackslash}m{8cm}            
    }
    \toprule
    \textbf{Feature Type} & \textbf{Index} & \textbf{Feature} & \textbf{Description} \\
    \midrule
    \multirow{13}{*}{\centering Net Vertex} & 1  & cell\_x & The x-coordinate of the cell instance on the device \\
    & 2  & cell\_y & The y-coordinate of the cell instance on the device \\
    & 3  & if LUT & Whether the instance is a lookup table  \\
    & 4  & if FF & Whether the instance is a flip-flop \\
    & 5  & if DSP & Whether the instance is a DSP module \\
    & 6  & if RAMB & Whether the instance is a block RAM \\
    & 7  & if MUX & Whether the instance is a multiplexer \\
    & 8  & if IO & Whether the instance is an I/O buffer \\
    & 9  & if ClockBuffer & Whether the instance is a clock buffer \\
    & 10 & if CARRY8 & Whether the instance is a carry chain \\
    & 11 & if Shifter & Whether the instance is a shifter \\
    & 12 & if LUTRAM & Whether the instance is a LUT-based RAM \\
    & 13 & if IN & Direction of signal propagation in the net relative to this instance \\
    \midrule
    \multirow{5}{*}{\centering Net Environment} & 14 & HPWL & Half-Perimeter Wire Length of the net \\
    & 15 & Width & Vertical width of the net \\
    & 16 & Length & Horizontal length of the net \\
    & 17 & Fanout & Fanout count of the net \\
    & 18 & Ave Routing Density & Average routing congestion of the net \\
    \midrule
    \multirow{9}{*}{\centering Pin Routing} & 19 & PinPairXDist & The x-direction distance between driver and load pin pairs \\
    & 20 & PinPairYDist & The y-direction distance between driver and load pin pairs \\
    & 21 & Net Index & Index of the load pin in the net \\
    & 22 & I/O Crossing & Whether the connection crosses an I/O module \\
    & 23 & DSP Crossing & Whether the connection crosses a DSP module \\
    & 24 & BRAM Crossing & Whether the connection crosses a BRAM module \\
    & 25 & Ave Pin Density & Average pin congestion between the driver and the load \\
    \bottomrule
    \end{tabular}
  \end{threeparttable}
  \label{datasets1}
\end{table*}

\subsection{\textbf{Data Preparation for Timing Analysis}}
\subsubsection{\textbf{Timing-Related Features Preprocessing}}\label{subsubsec:features}

The continuous expansion of FPGA design scales has led to interconnect delays being influenced by multiple nonlinear factors \cite{chang2000timing}.
To achieve accurate net delay prediction, TD-Placer constructs 25 key timing features based on circuit knowledge and parameter pruning experiments. 
As presented in Table~\ref{datasets1}, these features are systematically categorized into (a) net vertex features (characterizing pin locations and instance attributes), (b) net environment features (describing the operational context of pin nets), and (c) pin routing features (representing the routing environment between driver and load pin pairs). 
Notably, all net-intrinsic features except congestion metrics are directly extracted from the current placement state and original circuit design, ensuring the efficiency of TD-Placer.

The limited routing resources in FPGAs often force routing wires in congested areas to detour through non-congested regions, significantly impacting net delays \cite{xiong2024data}. 
To address this, TD-Placer adapts the congestion estimation method from RippleFPGA \cite{gengjie2018ripplefpga} by developing an average routing density metric to quantify localized congestion levels. 
Furthermore, we introduce an average pin density metric specifically between driver and load pins to more accurately characterize how routing congestion affects net delays.

\textbf{Average Routing Density.} The average routing density reflects the congestion level in the region traversed by a given net. 
In TD-Placer, the placement-target area is partitioned into uniform square grids, where the grid size is a constant determined experimentally. 
Each grid is treated as a global routing cell (g-cell), and the routing density of the $i$-th g-cell is computed as follows:

\begin{equation}
    con{g_i} = \sum\limits_{m \in {M_i}} {\frac{{NW(m) \cdot HPWL(m)}}{{{A_m}}}}
\end{equation}
where ${M_i}$ denotes the set of nets intersecting with the $i$-th g-cell, ${A_m}$ represents the total area of all g-cell covered by the net $m$, $NW$ is the weighting coefficients associated with pin count, and $HPWL$ corresponds to the HPWL estimation of the net $m$.

The average routing density of a net is defined as the mean density of all g-cells it traverses. 
As illustrated in Figure 5, the net connecting the driver and load pins has an average routing density equal to the mean value of the density within the the black shaded region, calculated as $(46 + 16 + 36 + 8 + 66 + 300) / 6  \approx 78.67$.

\textbf{Average Pin Density.} The average pin density reflects congestion between driver and load pins, providing a more refined and direct congestion metric for net delay. For any pin pair, their average pin density is defined as the number of utilized pins within their enclosing rectangular region.

\begin{table*}[!t]
  \centering
  \caption{\textsc{Resource Utilization of Benchmarks on VU095 Device}}
  \begin{threeparttable}
  \footnotesize
    \renewcommand{\arraystretch}{1.1}
    \begin{tabular}{
      >{\centering\arraybackslash}m{1.1cm}  
      >{\centering\arraybackslash}m{1.5cm}  
      >{\centering\arraybackslash}m{1.9cm}  
      >{\centering\arraybackslash}m{1.9cm}  
      >{\centering\arraybackslash}m{1.6cm}  
      >{\centering\arraybackslash}m{1.7cm}  
      >{\centering\arraybackslash}m{1.8cm}  
      >{\centering\arraybackslash}m{1.8cm}  
      >{\centering\arraybackslash}m{0.9cm}  
    }
    \toprule
    Benchmark & BLSTM\cite{rybalkin2017hardware} & DigitRecog\cite{zhou2018rosetta} & FaceDetect\cite{zhou2018rosetta} & SpooNN\cite{kara2018spoonn} & MemN2N\cite{sukhbaatar2015end} & MiniMap2\cite{guo2019hardware} & OpenPiton\cite{balkind2016openpiton} & VU095\\
    \midrule
    \#LUT     & 118,967 & 151,636 & 68,945 & 63,095 & 184,997 & 407,586 & 180,388 &537,600\\
    \#FF      & 54,690  & 105,580 & 56,987 & 70,987 & 84,694  & 252,624 & 111,966 &537,600\\
    \#CARRY   & 2,762   & 1,970   & 4,978  & 2,091  & 11,528  & 19,826  & 1,712 &33,600\\
    \#MUX     & 36,210  & 4,662   & 2,177  & 217   & 4,466   & 180    & 13,696 &201,600\\
    \#LUTRAM  & 1,147   & 251    & 255   & 251   & 3,500   & 251    & 752 &19,200\\
    \#DSP     & 258     & 1    & 101   & 165   & 312    & 528    & 58 &768\\
    \#BRAM    & 812    & 379    & 141   & 208   & 148    & 283    & 147 & 1,728\\
    \#Cell    & 215,101 & 265,775 & 134,450 & 137,937 & 289,721 & 681,889 & 309,145 & -\\ 
    \bottomrule
    \end{tabular}
  \end{threeparttable}
  \label{dataset2}
\end{table*}

\begin{figure}[!t] 
\centering
 \includegraphics[width=0.8\linewidth]{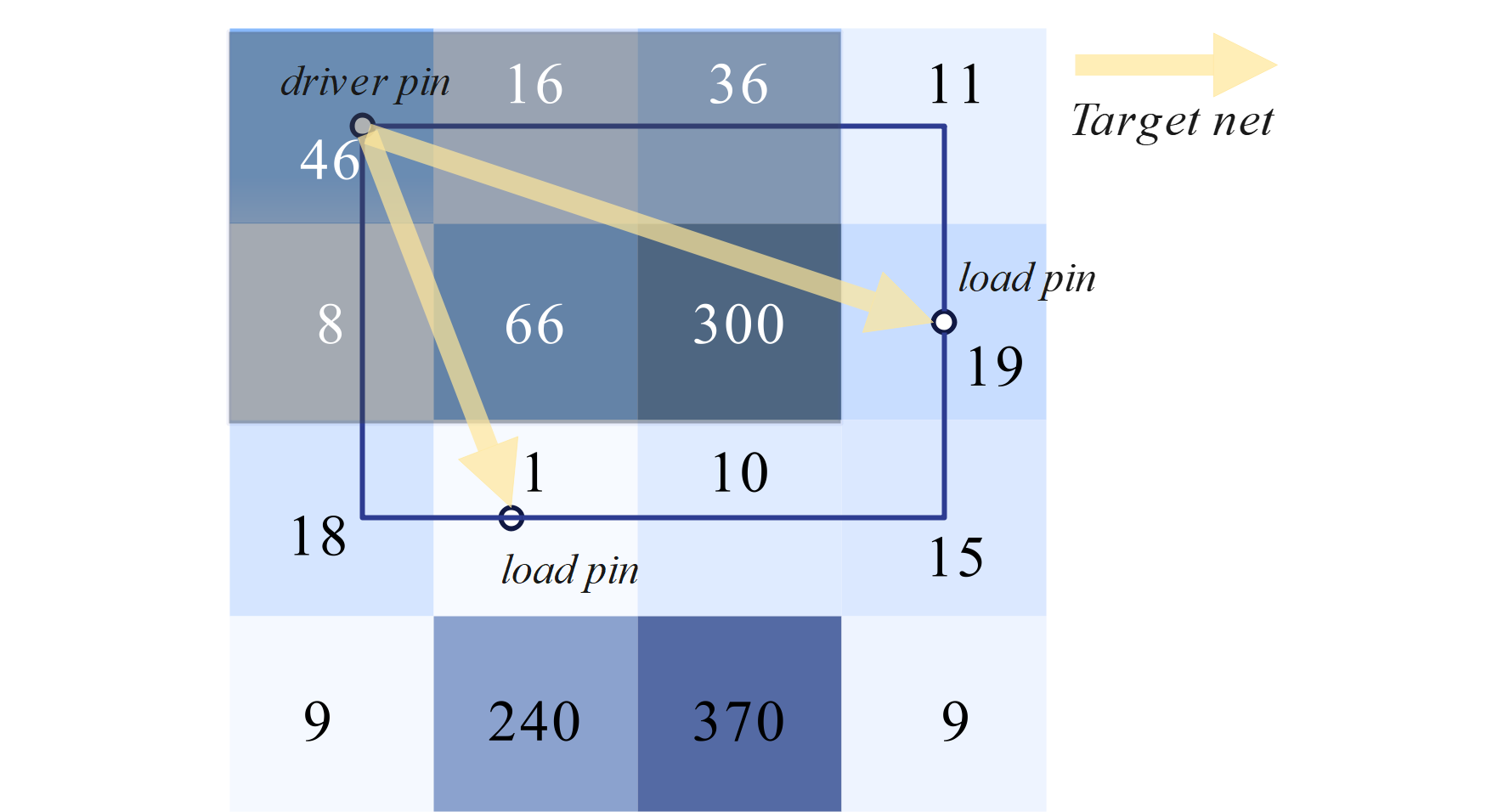}
\caption{The Average Routing Density of a net.}
\label{congfigure}
\end{figure}


\subsubsection{\textbf{Data Collection}} \label{datasettvt}

The ISPD 2016/2017 benchmarks \cite{yang2016routability,yang2017clock} have certain limitations: their randomly-generated netlists may produce unrealistic interconnects and unnecessary register duplication. Additionally, the absence of design hierarchy in these netlists results in an overly uniform distribution of critical timing paths, making them less representative of real-world scenarios\cite{liang2021amf}. To address this, we adopted seven open-source projects (Table \ref{dataset2}) as benchmarks to construct a dataset providing more realistic training and testing samples.

We extracted net vertex features from post-routing benchmarks using the commercial tool Vivado \cite{Xilinx2022}. For each driver-load pin pair within a net, we further extracted net environment features and pin routing features. All labels (net delays) were obtained using Vivado's built-in command \textit{get\_net\_delay}. By comprehensively traversing all nets across the seven benchmarks, we treated each pin pair as an independent sample in the dataset. However, the full sample size was prohibitively large. To ensure efficient timing analysis, we needed to build a lightweight deep learning model. Therefore, we adopted a balanced sampling strategy based on both the proportion of benchmark instances and the geometric distance between sample components to select representative samples. We divided the number of nets in the dataset into training, validation, and testing sets in a ratio of 7:1.5:1.5 and obtained 197178 training samples (44082 nets), along with 41734 validation samples (9447 nets) and 44737 test samples (9446 nets).

\subsection{\textbf{Net Information Extraction Component}}\label{sec:C}

\begin{figure*}[!t] 
\centering
 \includegraphics[width=0.85\linewidth]{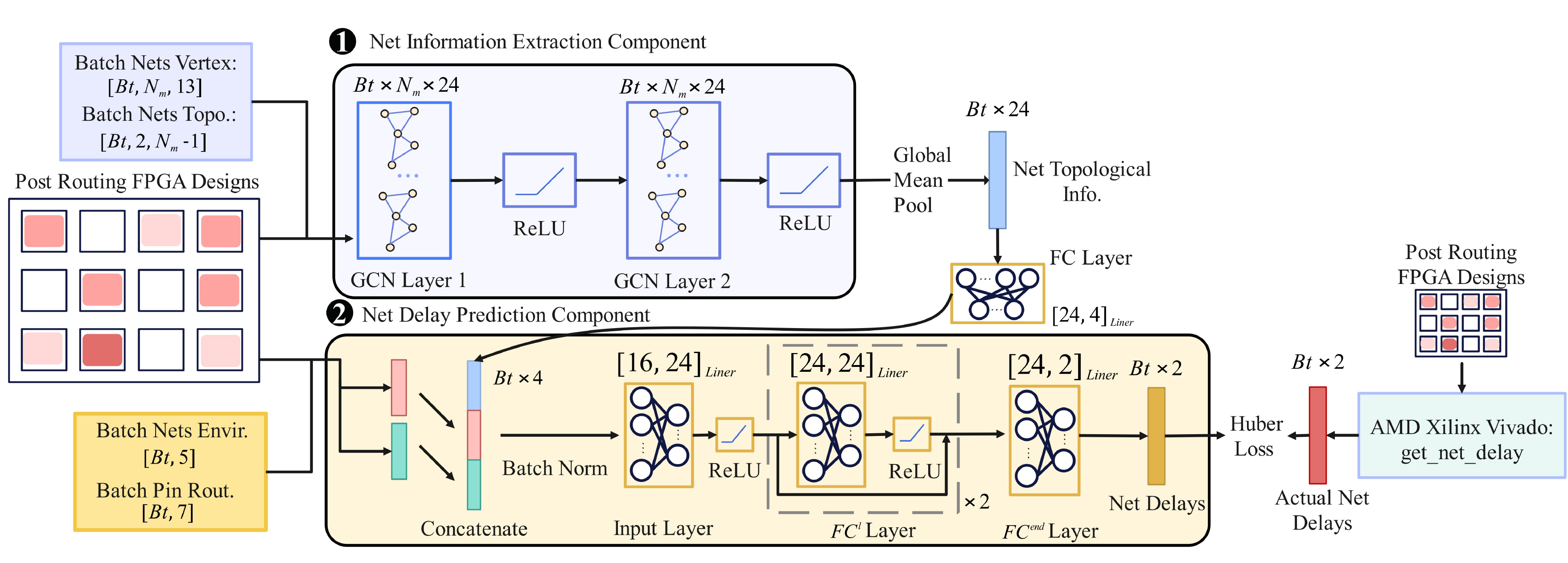}
\caption{Interaction and Architecture of Net Information Extraction and Net Delay Prediction Components.}
\label{NetPredict}
\caption*{\small \textit{The figure illustrates the batch training process of the network information extraction and delay prediction components, where $Bt$ denotes the batch size and $N_m$ represents the number of vertices in net $m$.}}
\end{figure*}

During FPGA routing, pins of the same net are interconnected through wires, forming a tree topology with the driver pin as the root and load pins as leaves. 
The net delay from the root to any leaf node is intrinsically linked to this tree's overall structure. 
To capture topological information, TD-Placer employs graph convolution mechanisms \cite{scarselli2008graph} to model load and crosstalk coupling effects in multi-driver scenarios. 
This method enables each vertex to aggregate feature information from its neighboring vertices during message passing. 

\textbf{Graph Construction for Nets:} 
For any given net $m$ composed of $N$ instance vertices and their driver-load relationships, TD-Placer constructs a corresponding directed graph $G(m) = {V, E, T}$, where $V$ denotes the vertex set, $E$ the set of directed edges, and $T$ the edge types.

\textbf{Vertex Set $V$:} 
For any given net, its instances vertices are represented by vertices $v$ belonging to the directed graph $G$'s vertex set $V$. 
Each vertex is initialized using the net vertex features defined in \ref{subsubsec:features}, and its embedding vector is iteratively updated during graph convolution.

\textbf{Edge Set $E$:} 
The edge set is constructed based on interconnects within the net. 
The existence of edge ${u_{ij}}$ $ \in $ $E$ indicates a direct connection between instance vertices ${v_i}$ and ${v_j}$.

\textbf{Edge Type Set $T$:} 
For each edge ${u_{ij}}$, its type ${t_{ij}}$ $ \in $ $T$ is defined by the driver-load relationship between instances. 
Four primary edge types $p \to p,p \to d,d \to d$ are established, where $p$ denotes the driver vertex, $d$ represents the driven vertex. 
To prevent vertex self-attributes from being disregarded during graph convolution, TD-Placer enforces self-loop connections for both driver and load vertices.

\textbf{Hierarchical Graph Convolution:} 
TD-Placer employ multi-layer graph convolution to jointly learn attributes of instance vertices and their interconnect topology. 
Through driver-load relationships, graph convolution aggregates and transforms features from adjacent instance vertices, which are then propagated to subsequent layers via the following implementation:

\begin{equation}
    v_i^{l + 1} = \sigma \left( {\left( {\sum\limits_{j \in P\left( { * ,i} \right)} {{w_{ji}}v_j^l} } \right){w^l}} \right)
\end{equation}
where $P\left( { * ,i} \right)$ denotes the set of vertices pointing to vertex ${v_i}$, $v_j^l$ represents the feature vector of the vertex after being updated at layer $l$, $v_i^{l + 1}$ indicates the feature vector after layer $l + 1$'s update, $\sigma $ is the nonlinear activation function, ${w^l}$ corresponds to the trainable parameter matrix at layer $l$, with ${w_{ji}}$ defined as:

\begin{equation}
    {w_{ji}} = \frac{1}{{\sqrt {\left( {\left| {{N_{in}}\left( i \right)} \right|\left( {\left| {{N_{out}}\left( j \right)} \right|} \right)} \right)} }}
\end{equation}
where ${N_{in}}(i)$ denotes the in-degree of vertex $i$, ${N_{out}}(j)$ represents the out-degree of vertex $j$.

After multi-layer graph convolution, each vertex embedding now encodes multi-hop neighborhood information. 
A global pooling operation is then applied to these vertex features to derive a unified graph-level representation capturing the complete topological information:

\begin{equation}
    y = \frac{1}{N}\sum\limits_{i = 1}^N {{v_i}}
\end{equation}
where $y$ is the feature vector embedding net topology information, $N$ is the number of component vertices in the current net, ${v_i}$ is the feature vector of the $i$-th vertex.

\subsection{\textbf{Net Delay Prediction Component}} \label{sec:D}

After processing by net information extraction component, TD-Placer obtains net topology information.
It then fuses this with net environment and pin routing features via a multi-layer perceptron (feature vector construction detailed in \ref{subsubsec:features}), enabling nonlinear modeling of net delay through multi-factor interactions. Figure \ref{NetPredict} shows the interaction between extraction and prediction components, along with internal model details.

Firstly, due to net topology dimensionality far exceeding that of the other two feature groups, this imbalance may hinder net prediction.
TD-Placer addresses this by passing topology features through a Fully Connected neural network for dimensionality reduction:

\begin{equation}
    y' = FC(y)
\end{equation}
where $y'$ denotes the dimensionality-reduced global feature vector of the net.

Next, TD-Placer concatenates the newly generated global net feature vector $y'$ with the other two feature groups:

\begin{equation}
    {y_{cat}} = \left[ {y',{y_{net}},{y_p}} \right]
\end{equation}
where ${y_{cat}}$ represents the concatenated global feature vector, ${y_{net}}$ denotes the feature vector of the net environment, ${{\rm{y}}_p}$ indicates the feature vector of the driver-load pin pair.



Following Batch Normalization, the feature vectors are then fed into a multi-layer feedforward neural network for feature fusion, with residual blocks incorporated to prevent gradient vanishing during training. 
Finally, a linear layer processes the fused features to get the final net delay $netDelay$:

\begin{equation}
  {\hat y^{l + 1}} = {\mathop{\rm Re}\nolimits} Lu(F{C^l}({\hat y^l})) + {\hat y^l}  
\end{equation}
\begin{equation}
  netDelay = F{C^{end}}({\hat y^{end}})
\end{equation}
where ${\hat y^{l + 1}}$ represents the composite feature vector at layer $l + 1$, $F{C^l}$ denotes the weight matrix of the feedforward neural network at layer $l$, $F{C^{end}}$ indicates the weight matrix of the linear layer, ${\hat y^{end}}$ corresponds to the feature vector after feedforward neural network processing.

TD-Placer employs the Smooth Mean Absolute Error loss function \cite{gokcesu2021generalized} for loss minimization. Since net delays are influenced by complex factors (e.g., FPGA architecture, temperature), extreme delay values (outliers) inevitably occur. These outliers may distort the model’s judgment and trap it in local optima. Huber Loss is adopted to mitigate the impact of extreme values, ensuring robust model performance.

\begin{equation}
    \scalebox{0.85}{$
{L_\delta }\left( {\overline y ,netDelay} \right) = \left\{ 
\begin{array}{@{}l@{\quad}l@{}}
\scriptstyle \frac{1}{2}{(\bar y - netDelay)}^2, & \scriptstyle \text{if } |\bar y - netDelay| \le \delta \\
\scriptstyle \delta |\bar y - netDelay| - \frac{1}{2}\delta^2, & \scriptstyle \text{if } |\bar y - netDelay| > \delta 
\end{array}
\right\}
$}
\end{equation}
where $\bar y$ denotes the ground truth net delay, $\delta $ represents the tunable threshold parameter controlling outlier boundaries.

\subsection{\textbf{Global Placement Component}}

This section outlines the full global placement workflow, highlighting TD-Placer’s lightweight integration of the precise timing model introduced in Sections~\ref{sec:C} and~\ref{sec:D} for static timing analysis, as well as its fine-grained timing-aware weighting method to effectively optimize circuit timing.

\begin{table*}[!t]
  \centering
  \caption{\textsc{Comparison of Placement Runtime (s)}}
  \begin{threeparttable}
    \renewcommand{\arraystretch}{1.1} 
    \begin{tabular}{
      >{\centering\arraybackslash}m{4cm}  
      >{\centering\arraybackslash}m{1.0cm}  
      >{\centering\arraybackslash}m{1.0cm}  
      >{\centering\arraybackslash}m{1.0cm}  
      >{\centering\arraybackslash}m{1.0cm}  
      >{\centering\arraybackslash}m{1.0cm}  
      >{\centering\arraybackslash}m{1.0cm}  
      >{\centering\arraybackslash}m{1.0cm}  
      >{\centering\arraybackslash}m{1.0cm}  
    }
    \toprule
    Benchmark  & BLSTM & DigitRecog & FaceDetect  & SpooNN & MemN2N & MiniMap2 & OpenPiton & Average \\
\midrule
AMF-Placer\cite{liang2024amf} & 272 & 366 & 159 &	142  & 366 & 543 & 346 & 0.92\\ 
\midrule
TD-Placer (\textit{no lightweight alg.}) & 408	& 512 & 170 &	182 & 387 & 876 & 547 & 1.29\\ 
\midrule
TD-Placer& 284	& 408 & 162 &	160 & 372 & 617 & 384 & 1\\ 
\bottomrule
\end{tabular}%
  \end{threeparttable}
  \label{expriment0}%
\end{table*}

\begin{algorithm}[!t]
\caption{ \textbf{The Process of Static Timing Analysis}}
\label{alg_sta}
\SetKwInOut{Input}{Input}
\SetKwInOut{Output}{Output}
\SetKwInOut{Require}{Require}
\SetKwFunction{FBuildPrompt}{BuildPrompt}
\SetKwFunction{FGenerate}{Generate}  
\SetKwFunction{FCheckEquiv}{CheckEquivalence}
\SetKwProg{Fn}{Function}{:}{}

\Require{timing graph $G$, logic delay lookup table $T_l$, weight of graph convolution $W_t$, weight of mlp $W_m$, batch size $bt$}
\Output{timing graph $G'$, critical path $P$}

\BlankLine
Initialize $i \leftarrow 0$\;
Initialize Net vertex Set $\mathcal{A}$, Net Environment Array $\mathcal{B}$, Pin Routing Array $\mathcal{C}$, Net Delays $\mathcal{D}$ ;
\vspace{2pt} 

$ G' \leftarrow updateVertexWeights(G, T_l)$

\vspace{1pt} 
\ParFor{$i = 1$ \KwTo $getEdgesSize(G)$}{

    $net \leftarrow getNetFromGraph(G,i)$\;
    \vspace{1pt} 

    $setFeatures(net,i,\mathcal{A},\mathcal{B},\mathcal{C})$\;  
    \vspace{1pt}
} 

$\mathcal{A'} \leftarrow batchGetNetTopologyInfo(\mathcal{A},W_t,bt)$\;
\vspace{1pt} 
$\mathcal{D} \leftarrow batchGetNetDelay(\mathcal{A'},W_m,bt)$\;
\vspace{1pt} 
\ParFor{$i = 1$ \KwTo $getEdgesSize(G)$}{

    $updateGraphArcWeights(G',\mathcal{D},i)$\;
    \vspace{1pt} 
} 

$ P \leftarrow forwardPropagation(G')$
\vspace{1pt}

$backwardPropagation(G')$

\Return $G',P$

\end{algorithm}

\subsubsection{\textbf{Static Timing Analysis}}
To reduce the computational complexity of deep learning, TD-Placer employs a lightweight net delay prediction algorithm. 
Simultaneously, it adopts a hierarchical parallelization strategy to efficiently compute pin-level timing slack through forward and backward propagation. 
\paragraph{\textbf{Proposed Lightweight Algorithm}}
During the initial global placement phase, the hierarchical topology of the design netlist is abstracted into a directed acyclic timing graph. 
The vertex weights in this graph are initialized using a pre-constructed logic delay lookup table. As the physical locations of instances are dynamically adjusted throughout the iterative placement process, net delays are recomputed to update the edge weights in the timing graph.

To handle the one-to-many driver-load pin pairs within each net and avoid redundant computations in end-to-end net delay prediction, TD-Placer employs a weight-decoupled architecture that separates graph convolution (for topology extraction) and MLP (for delay regression), with an optimized inference flow (Algorithm \ref{alg_sta}, lines 1–10).
Specifically, the method begins by traversing all timing arcs in the timing graph to update net vertex features, pin routing features, and net environment features on a per-net basis (Lines 4-6 in Algorithm 1).
Subsequently, the net information extraction component performs batched processing using GPU acceleration and conducts convolutional inference to efficiently acquire topological information of the nets (Line 7 in Algorithm 1).
The net delay prediction component then concatenates the obtained topology information with corresponding pin routing features and net environment features.
These samples are packaged into small batches and fed into the GPU for inference, ultimately yielding net delay predictions (Line 8 in Algorithm 1).
Finally, the method performs another parallel traversal of all arcs in the timing graph to update their timing weights (Lines 9-10 in Algorithm 1).
This algorithm demonstrates significant reduction in computational overhead for timing analysis, particularly when handling large-scale netlists.
Furthermore, TD-Placer parallelizes the feature migration process from CPU to GPU, thereby achieving additional improvements in overall computational efficiency.

We compare the runtime of TD-Placer with and without the lightweight algorithm integrated into placement. As shown in Table~\ref{expriment0}, placement time is reduced by an average of 29\% with the lightweight algorithm. TD-Placer incurs 8\% more runtime than AMF-Placer due to deep learning overhead. Incorporating more features increases inference parameters, which may be mitigated by larger batch sizes and greater GPU memory.

\paragraph{\textbf{Parallelized Propagation}} A path delay typically comprises both logic and net delays.
Upon completing net delay prediction, both are stored in the directed acyclic timing graph as vertex and arc weights, respectively.
TD-Placer then performs forward and backward propagation on the graph following OpenTimer's \cite{huang2015opentimer} timing analysis methodology (Algorithm~\ref{alg_sta}, lines 11–12) to determine slack for each arc.
Since vertices at the same timing level are mutually independent, TD-Placer parallelizes the process to reduce runtime.
Starting from the first level (outputs of source registers), TD-Placer conducts levelized forward propagation to determine the latest arrival time for each instance:
\begin{equation}
T_{arr}(v_i) = \mathop{\max}\limits_{\forall v_j \in fanin(v_i)} \left( T_{arr}(v_j) + logic(v_j) + netD(u_{ji}) \right)
\end{equation}
where $T_{arr}(v_i)$ denotes the latest signal arrival time at the $i$-th instance, $logic(v_j)$ is the logic delay of the $j$-th instance, $netD(u_{ji})$ is the net delay of edge $u_{ji}$, and $fanin(v_i)$ specifies the set of fan-in instances for the $i$-th instance.

Subsequently, TD-Placer performs backward propagation starting from the final level in a levelized manner to determine the earliest required arrival time for each instance:
\begin{equation}
    {T_{req}}({v_i}) = \mathop {\max }\limits_{\forall {v_j} \in fanout({v_i})} ({T_{req}}({v_j}) - logic({v_j}) - netD({u_{ij}}))
\end{equation}
where ${T_{req}}({v_i})$ denotes the latest required arrival time at the $i$-th instance, $fanout$ represents the set of fan-out instances for instance $i$.

Finally, TD-Placer obtains the timing slack $Slac{k_{ij}}$ between driver instance $i$ and load instance $j$, calculated as:
\begin{equation}
    Slac{k_{ij}} = {T_{req}}(j) - {T_{arr}}(i) - netD({u_{ij}})
    \label{eq:slack}
\end{equation}

While our accurate timing model is time-consuming, the lightweight algorithm mitigates computational costs, and the resulting precise slack offers effective guidance for subsequent timing optimization.


\subsubsection{\textbf{Timing-Driven Quadratic Placement}}

As noted in Section \ref{sec:TGP}, wirelength-driven placer minimizes weighted wirelength\cite{liang2021amf,wuxi2019new}, while timing-driven placement adds timing terms to optimize TNS and WNS. TD-Placer introduces a pin-to-pin weighting scheme and proposes a global timing and logic-depth-aware strategy to optimize high-risk timing arcs.

\paragraph{\textbf{Timing-arc Based Quadratic Placement Formulation}}

For high-fanout FPGA designs, applying a uniform worst-case slack-based weight to an entire net can leave timing arcs with positive slack unconstrained, significantly degrading overall TNS \cite{xiong2024data}. 
To address this, TD-Placer directly constructs fine-grained pin-to-pin timing loss. 
In addition, following the original implementation in \cite{liang2024amf}, for long timing paths that span a single FPGA clock region, TD-Placer aims to place all instances on the path within the same clock region. 
To achieve this, it generates pseudo nets at the horizontal center of the clock region containing more than 50\% of the instances on the path, and connect them to the corresponding instances. 
This guides instances toward the target clock region during optimization and enables vertical placement alignment to alleviate congestion. Based on these considerations, the quadratic cost function for placement is formulated as:
\begin{subequations}
\label{eq:main}  
\begin{align}
\mathop{\min}\limits_{x,y} \ & (1 - \lambda)(WL + WD) + \lambda WT \label{eq:main_obj} \\
\text{with:} \quad 
WL &= \sum_{m \in Q} W_m \label{eq:wl} \\
WD &= \sum_{m_p \in Q_p} w_{m_p} W_{m_p}(x, y) \label{eq:wd} \\
WT &= \sum_{u_t \in Q_t} w_{u_t} W_{u_t}(x, y) + \sum_{u_b \in Q_b} w_{u_b} W_{u_b}(x, y) \label{eq:wt}
\end{align}
\end{subequations}
where:
\begin{itemize}
    \item $\lambda $ balances the wirelength and timing optimization terms during global placement.
    \item ${Q_b}$ denotes the clock-region-aware pseudo nets set, ${W_{u_b}}$ represents their HPWL-based approximations, and ${w_{u_b}}$ is the corresponding weight applied to the pseudo nets\cite{liang2024amf}.
    \item ${Q_t}$ is the set of timing arcs, and ${w_{{u_t}}}$ is the timing slack-related weight. The slack is defined in Equation \ref{eq:slack}. And ${W_{{u_t}}}$ is the HPWL approximation of timing arcs. 
\end{itemize}

\paragraph{\textbf{Depth and Global Timing-Aware Weighting Scheme}}
In prior timing-driven placer \cite{xiong2024data}, ${w_{{u_t}}}$ is defined as:
    \begin{equation}
    {w_{{u_t}}} = \left\{ {\begin{array}{*{20}{c}}
0&{slack({u_t}) \ge 0}\\
{{{\left( {1 - \frac{{slack({u_t})}}{{C{l_{require}}}}} \right)}^\alpha }}&{slack({u_t}) < 0}
\end{array}} \right.
\end{equation}
where $C{l_{require}}$ is the target delay value, and $\alpha $ is a constant determined empirically.

\begin{figure}[!t] 
\centering
 \includegraphics[width=1.0\linewidth]{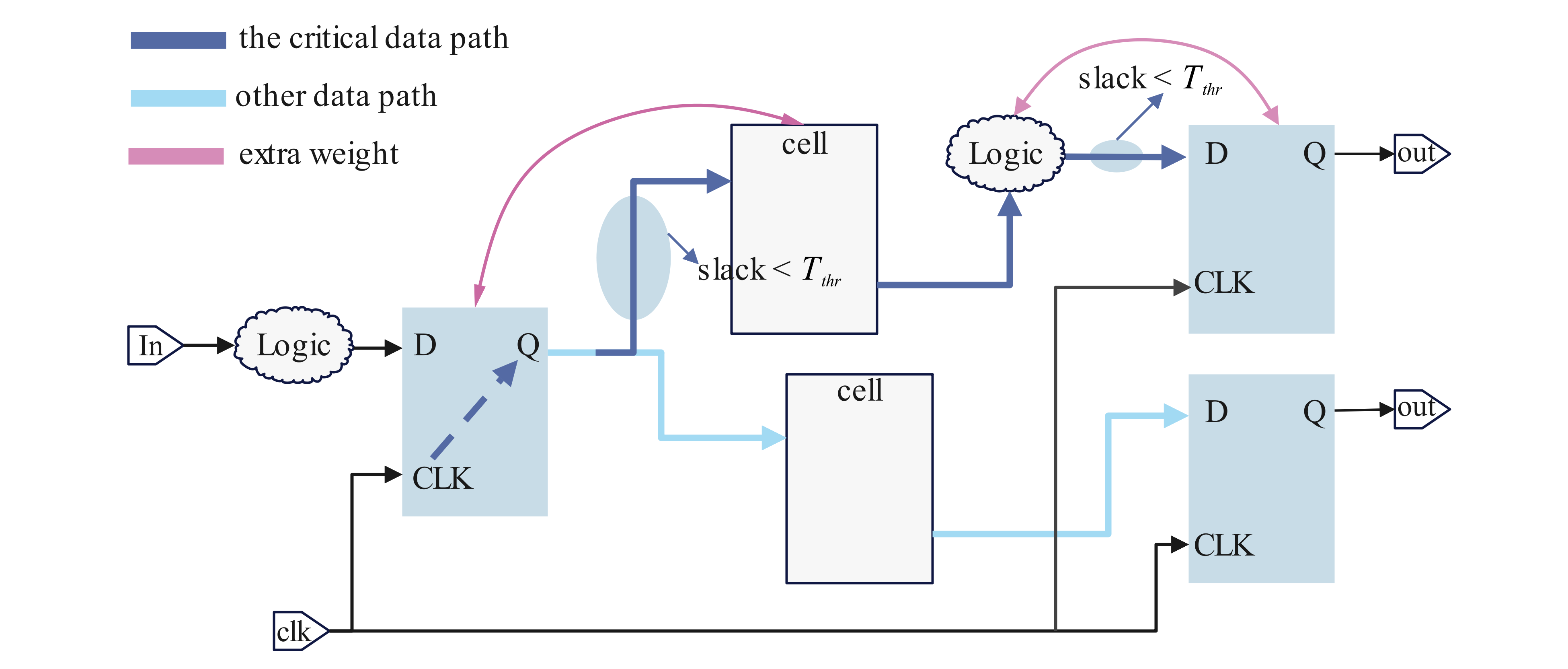}
\caption{The illustration of optimizations for the critical path.}
\label{congfigure}
\end{figure}

While inspired by \cite{liang2024amf}, our method differs in that TD-Placer directly applies additional slack-related weighting coefficients to timing arcs with slack below a predefined threshold, thereby enhancing their optimization strength.
Furthermore, as show in Figure \ref{congfigure}, if the arc lies on a critical timing path, a small incremental weight is applied to ensure that, within similar slack ranges, the critical path receive higher optimization priority, thereby enabling the delay of the critical path to decrease more steadily. 
To further refine the optimization strategy for the critical path, TD-Placer introduces an internal prioritization mechanism among timing arcs on the path: arcs farther from the endpoint register are assigned higher weights. 
This design is based on the observation that optimizing early-stage arcs not only improves local timing but also has a significant cascading effect on the timing of downstream instances, thereby maximizing overall timing improvement.

Therefore, in TD-Placer, the final weighting coefficient 
${w_{{u_t}}}$ for each timing arc is defined as follows:

\begin{equation}
    {w_{{u_t}}} = {\left( {1 - \frac{{slack({u_t})}}{{C{l_{require}}}}} \right)^{\alpha  + \frac{{\beta slack({u_t})}}{{{T_{thr}}}} + \varsigma }}
\end{equation}
where $\beta$ is a constant, ${T_{thr}}$ is the percentile value used to evaluate the current global timing quality, and $\varsigma$ is the additional weight assigned to identified critical timing paths, defined as follows:
\begin{equation}
    \varsigma  = \left\{ {\begin{array}{*{20}{c}}
0&{{u_t} \notin pat{h_{critical}}}\\
{\gamma \frac{{{c_{forward}}}}{{{c_{\max }}}}}&{{u_t} \in pat{h_{critical}}}
\end{array}} \right.
\end{equation}where ${c_{forward}}$ is the distance from the current load pin to the endpoint of the critical timing path, ${c_{\max }}$ is the total length of the critical path, $pat{h_{critical}}$ denotes the critical timing path, and $\gamma $ is a small constant determined experimentally.

\begin{table*}[!t]
  \centering
  \caption{\textsc{Comparison of Placement WNS, TNS, and Post-route CPD (ns) Using Vivado 2020 and 2021}}
  \begin{threeparttable}
  \scriptsize
    \renewcommand{\arraystretch}{0.95} 
    \begin{tabular}{
      >{\centering\arraybackslash}m{1.2cm}  
      >{\centering\arraybackslash}m{1.9cm}  
      >{\centering\arraybackslash}m{1.0cm}  
      >{\centering\arraybackslash}m{1.0cm}  
      >{\centering\arraybackslash}m{1.0cm}  
      >{\centering\arraybackslash}m{1.0cm}  
      >{\centering\arraybackslash}m{1.0cm}  
      >{\centering\arraybackslash}m{1.0cm}  
      >{\centering\arraybackslash}m{1.0cm}  
      >{\centering\arraybackslash}m{1.0cm}  
    }
    \toprule
    EVA & Place-Route  & BLSTM & DigitRecog & FaceDetect  & SpooNN & MemN2N & MiniMap2 & OpenPiton & Average \\
    \midrule
    \addlinespace
    \multirow{6}{*}{WNS} 
& V2020 \cite{Xilinx2020} & -0.562 & \textbf{-2.564} & -0.243 & -0.779 & \textbf{-0.669} & 0.037 &
\textbf{-2.159} & \textbf{0.85}\\
& AMF-V2020 \cite{liang2024amf}& -0.389 & -3.595 & -0.384 & -0.491 & -1.601 & 0.049 &
-2.310 & 1.07 \\
& TD-Placer-V2020 & \textbf{-0.33}	& -3.486 &	\textbf{0.126} &	\textbf{-0.489} & -1.581 &	\textbf{0.068} &  -2.49 & 1 \\
\addlinespace
\cline{2-10}
\addlinespace
& V2021 \cite{Xilinx2021} & -0.668 & \textbf{-3.249} & -0.264 & -0.836 & \textbf{-0.732} & \textbf{0.07} &
\textbf{-2.436} & \textbf{0.88} \\
& AMF-V2021 \cite{liang2024amf} & -0.508 & -4.373 & -0.445 & -0.537 & -1.665 & 0.001 & -2.556 & 1.12 \\
& TD-Placer-V2021 & \textbf{-0.356} & -3.98 & \textbf{0.012} & \textbf{-0.517} &	-1.763 &	0.017 & -2.517 & 1 \\
\addlinespace
\midrule
\addlinespace
\multirow{6}{*}{TNS} 
& V2020 \cite{Xilinx2020} & -18 & \textbf{-39689} & -1.865 & -17 & \textbf{-485} & 0 &
\textbf{-13493} & - \\
& AMF-V2020 \cite{liang2024amf}& -18 & -54278 & -0.413 & \textbf{-9}& -1269 & 0 &
-27189 & - \\
& TD-Placer-V2020 & \textbf{-10}	& -53714 &	\textbf{0} &	-14 & -1214 &	\textbf{0} &  -21595 & - \\
\addlinespace
\cline{2-10}
\addlinespace
& V2021 \cite{Xilinx2021} & -54	& \textbf{-46804}	& -0.273	&-18& \textbf{-352}	& 0	& \textbf{-9343} & - \\

& AMF-V2021 \cite{liang2024amf} & -21	& -64569	& -0.574	& -18	& -1676	& 0	& -28630 & - \\
& TD-Placer-V2021 & \textbf{-14}& -61372	& \textbf{0}	& \textbf{-14}	& -1624	& \textbf{0}	& -19775& - \\
\addlinespace
\midrule
\addlinespace

\multirow{12}{*}{CPD}
\multirow{2}{*}
& V2020 \cite{Xilinx2020} & 8.56 & \textbf{10.61} & 15.24 & 8.79 & \textbf{10.67} & 7.96 & 12.17 & - \\
& Rnorm & 1.1 & \textbf{0.95} & 1.02 & 1.08 & \textbf{0.95} & 1.02 & 1.0 & 1.02\\
\multirow{2}{*}
& AMF-V2020 \cite{liang2024amf}& 8.40 & 11.64 & 15.39 & 8.50 & 11.60 & 7.96 & 12.38 & - \\
& Rnorm  & 1.08 & 1.05 & 1.03 & 1.05 & 1.03 & 1.02 & 1.02 & 1.04 \\
\multirow{2}{*}
& TD-Placer-V2020 & \textbf{7.79} & 11.11 & \textbf{14.96} & \textbf{8.144} &  11.24 & \textbf{7.782} & \textbf{12.14} & - \\
 & Rnorm & \textbf{1} & 1 & \textbf{1} & \textbf{1} & 1 & \textbf{1} & \textbf{1} & \textbf{1} \\
\addlinespace
\cline{2-10}
\addlinespace
\multirow{2}{*}
& V2021 \cite{Xilinx2021} & 8.67 & \textbf{11.29} & 15.27 & 8.85 & \textbf{10.73} & 7.94 & 12.44 & - \\
& Rnorm & 1.15 & \textbf{0.93} & 1.02 & 1.1 & \textbf{0.93} & 1.08 & 1.01 & 1.03\\
\multirow{2}{*}
& AMF-V2021 \cite{liang2024amf} & 8.51 & 12.41 & 15.45 & 8.55 & 11.66 & 8.01 & 12.56 & - \\
& Rnorm & 1.13 & 1.03 & 1.03 & 1.06 & 1.01 & 1.1 & 1.02 & 1.05 \\
\multirow{2}{*}
& TD-Placer-V2021 & \textbf{7.56} & 12.1 & \textbf{14.98}	& \textbf{8.05}	& 11.55& \textbf{7.34}	& \textbf{12.31}	& - \\
& Rnorm & \textbf{1} & \textbf{1} & \textbf{1} & 1 & 1 & \textbf{1} & \textbf{1} & \textbf{1} \\

    \bottomrule
    \end{tabular}%

  \end{threeparttable}
  \label{expriment1.0}%
\end{table*}

\subsubsection{\textbf{Rough Legalization}}

In FPGAs, resources are location-constrained, requiring instance spreading after each placement iteration. TD-Placer adopts the method from \cite{liang2021amf}, dividing the area into grids, detecting overflows, and redistributing instances to nearby grids until resolved. After spreading, virtual pseudo nets are created based on new instance positions and used as legalization terms during quadratic placement (as shown in Equation \ref{eq:wd}) to control instance density.

\subsubsection{\textbf{Resource Demand/Supply Adjustment}} 
After each rough legalization step, to allow direct legalization and avoid routing congestion, it is necessary to adjust instance resource demand and available grid resources.
TD-Placer adopts the method from \cite{wuxi2019new}, dynamically adjusting the resource demand of corresponding instance types when LUT or FF positions change.
This method is extended to handle clock constraints by checking for clock overflow: if a half-column clock net in a region exceeds 80\% capacity, the FF resource demand for that clock net increases.
To further ensure routability, TD-Placer applies the congestion estimation approach from \cite{zhuo2006congestion}, dynamically adjusting available grid resources based on congestion levels and increasing cell spacing in congested regions during later iterations.

\section{\textbf{EVALUATION}}

\subsection{Experimental Setup}
\textbf{Implementation details.} TD-Placer is implemented in C++ and Python. Net extraction and delay prediction components are trained synchronously in Python and integrated into the C++ placement flow via model tracing. 

TD-Placer targets the Xilinx UltraScale VU095 FPGA device, but the related techniques can be easily adapted to other devices using general data preprocessing scripts provided by us. The benchmark tests used in the experiments (as shown in Table \ref{dataset2}) are all applicable to the VU095 device.


\textbf{Baselines.} Existing state-of-the-art timing-driven placers include methods proposed by Xiong \cite{xiong2024data}, Mai et al. \cite{mai2023openparf}, and AMF-Placer. We adopt AMF-Placer as our primary open-source baseline, as Xiong’s method does not support key FPGA primitives widely used in practical designs (e.g., MUX and CARRY8), and Mai et al.’s open-source library lacks essential timing analysis modules. For comprehensive evaluation, we also include five versions of the commercial placer AMD Xilinx Vivado, spanning from 2020.2 to 2024.2.

\textbf{Running Platform.} The experimental environment is an Ubuntu 22.04 server, with hardware configurations including an NVIDIA A6000 GPU, a 10-core 4.8GHz CPU (with Turbo Boost support), and 192GB of RAM.

\subsection{\textbf{Comparisons to state-of-the-art placers}}

\textbf{Motivation: }In this section, we evaluate the placement performance of TD-Placer on seven benchmarks (as shown in Table \ref{dataset2}) to verify its effectiveness in real-world scenarios. We integrate TD-Placer into the infrastructure of AMF-Placer \cite{liang2024amf}. Placement results on seven benchmarks are routed in Vivado, where post-routing CPD and WNS are extracted using the built-in report\_timing\_summary for evaluation.

\textbf{Results: }The performance of TD-Placer on seven benchmark test sets is shown in Tables \ref{expriment1.0} and \ref{expriment1.1}, where "V" represents Vivado, "AMF" represents AMF-placer, and "-" indicates a combination. For example, AMF-V2020 denotes the result of using AMF-placer for placement and Vivado 2020 for routing. Overall, TD-Placer shows a significant advantage over AMF-placer across all three metrics. On the seven benchmark tests with five corresponding Vivado routing versions, the CPD metric improved by an average of 4\%, 5\%, 4\%, 4\%, and 4\%, respectively. WNS metric improved by 7\%, 12\%, 6\%, 10\%, and 4\%. Additionally, TNS shows notable improvements in most scenarios. Specifically, in the Vivado 2020, 2021, and 2022 routing versions, FaceDetect transitioned from negative slack to positive slack. BLSTM, DigitRecognition, and Minimap2 demonstrated significant improvements across all Vivado routing versions, while MemN2N, SpooNN, and OpenPiton outperformed AMF-placer in most Vivado routing versions. Compared to the commercial tool Vivado, TD-Placer improved the CPD metric by an average of 2\% and 3\% for the 2020 and 2021 versions. For the 2022, 2023, and 2024 versions, the average degradation was only 1\%, 1\%, and 4\%, respectively, highlighting the excellent performance of our placer in timing optimization.

\begin{figure}[htbp]
    \centering
    \subfloat[\footnotesize \textnormal{TD-Placer}]{\includegraphics[width=0.22\columnwidth, keepaspectratio]{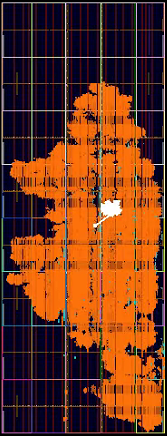}\label{fig:sub1}}
    \hspace{0.02\linewidth}
    \subfloat[\footnotesize \textnormal{AMF-Placer}]{\includegraphics[width=0.22\columnwidth, keepaspectratio]{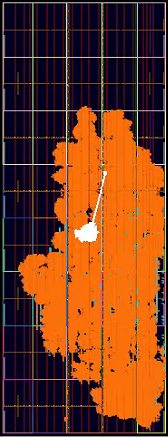}\label{fig:sub2}}
    \caption{Comparison between TD-Placer and AMF-Placer placement results on digitRecognition\cite{zhou2018rosetta}.}
    \label{placement_results}
\end{figure}

Compared to the baseline method, AMF-placer, TD-Placer demonstrates significant improvements in performance. Its advantages primarily stem from two key aspects:
\begin{itemize}
\item \textbf{Equipped with a precise timing model:}
FPGA net delay is influenced by multiple nonlinear factors, and load and crosstalk coupling effects in multi-pin driving scenarios can significantly impact delay. TD-Placer’s timing model considers diverse features and leverages global net representations, learning from large-scale samples to achieve accurate and generalizable delay prediction.

\item \textbf{Considering global timing and fine-grained critical paths weighting:}
TD-Placer dynamically assesses global timing by assigning extra weights to high-risk arcs, promoting optimization in critical regions. It further prioritizes critical paths based on logic depth to ensure stable delay reduction during iteration. As shown in Figure \ref{placement_results}, our placement on digitRecognition benchmark achieves a significantly shorter critical path than AMF-Placer.
\end{itemize}

\begin{table*}[!t]
  \centering
  \caption{ \textsc{Comparison of Placement WNS, TNS, and Post-route CPD (ns) Using Vivado 2022–2024}}
  \begin{threeparttable}
  \scriptsize
    \renewcommand{\arraystretch}{0.95} 
    \begin{tabular}{
      >{\centering\arraybackslash}m{1.2cm}  
      >{\centering\arraybackslash}m{1.9cm}  
      >{\centering\arraybackslash}m{1.0cm}  
      >{\centering\arraybackslash}m{1.0cm}  
      >{\centering\arraybackslash}m{1.0cm}  
      >{\centering\arraybackslash}m{1.0cm}  
      >{\centering\arraybackslash}m{1.0cm}  
      >{\centering\arraybackslash}m{1.0cm}  
      >{\centering\arraybackslash}m{1.0cm}  
      >{\centering\arraybackslash}m{1.0cm}  
    }
    \toprule
    EVA & Place-Route  & BLSTM & DigitRecog & FaceDetect  & SpooNN & MemN2N & MiniMap2 & OpenPiton & Average \\
    \midrule
    \addlinespace
    \multirow{9}{*}{WNS} 
& V2022 \cite{Xilinx2022}& -0.627 & -4.516 & 0.252 & \textbf{-0.349} & \textbf{-0.842}	& 0.009	& \textbf{-2.054} & \textbf{0.91}\\
& AMF-V2022 \cite{liang2024amf} & -0.5 & -4.62 & 0.37 & -0.481 & -1.876 & -0.001 & -2.385 & 1.06\\
& TD-Placer-V2022 & \textbf{-0.462} & \textbf{-3.927} & \textbf{0.407} & -0.681 & -1.922	& \textbf{0.024}	& -2.376 & 1 \\

\addlinespace
\cline{2-10}
\addlinespace

& V2023 \cite{Xilinx2023}& -0.972 & \textbf{-3.642} &	0.12 &	\textbf{-0.391} & \textbf{-0.698} & 0.048 & \textbf{-1.946} & \textbf{0.82} \\
& AMF-V2023 \cite{liang2024amf} & -0.489 & -4.862 &	0.085 &	-0.643 & -1.862 & -0.001 & -2.345 & 1.1\\
& TD-Placer-V2023 & \textbf{-0.435} &-4.216 & \textbf{0.216} & -0.703 & -1.812 & \textbf{0.067} & -2.297 & 1 \\

\addlinespace
\cline{2-10}
\addlinespace

& V2024 \cite{Xilinx2024}& -0.546 & \textbf{-2.917} & 0.086 & \textbf{-0.396} & \textbf{-0.648} & \textbf{0.046}	& \textbf{-2.182}  & \textbf{0.74}\\
& AMF-V2024 \cite{liang2024amf} & -0.387 & -4.275 & 0.296 & -0.672 & -1.684	& 0.024	& -2.47 & 1.04\\
& TD-Placer-V2024 & \textbf{-0.345} & -3.957 & \textbf{0.356} & -0.652 & -1.923	 & 0.029 &  -2.38 & 1 \\

\addlinespace
\midrule
\addlinespace
    \multirow{9}{*}{TNS} 
& V2022 \cite{Xilinx2022}& -11	&-69667	&0	&\textbf{-6}	&\textbf{-501}	&0	&\textbf{-6807}& - \\
& AMF-V2022 \cite{liang2024amf} & -27	&-66195	&0	&-20	&-1610	&0	&-26799 & - \\
& TD-Placer-V2022 & \textbf{-9}	& \textbf{-62801}	& \textbf{0}	&-14	&-1712	&\textbf{0}	&-20922 & -
 \\

\addlinespace
\cline{2-10}
\addlinespace

& V2023 \cite{Xilinx2023}& -44	&\textbf{-50880}	&0	&\textbf{-6}	&\textbf{-309}	&0&	\textbf{-11251}& - \\
& AMF-V2023 \cite{liang2024amf} & -23&	-64726&	0	&-18&	-1527&	-0.001&	-26994& -\\
& TD-Placer-V2023 & \textbf{-17}&	-61251	&\textbf{0}	&-12	&-1745	&\textbf{0}	&-19690 & -
 \\

\addlinespace
\cline{2-10}
\addlinespace

& V2024 \cite{Xilinx2024}& -13&	\textbf{-44019}&	0&	\textbf{-7}&	\textbf{-271}&	0	&\textbf{-9680} & -
\\
& AMF-V2024 \cite{liang2024amf} & -19&	-58803&	0&	-19&	-1649&	0&	-27530& -
\\
& TD-Placer-V2024 & \textbf{-8}	&-61307	&\textbf{0}	&-12	&-1522&	\textbf{0}	&-21363& -
 \\

\addlinespace
\midrule
    \addlinespace

\multirow{18}{*}{CPD}
\multirow{2}{*}
& V2022 \cite{Xilinx2022}& 8.07 &	12.38 &	17.4 &	\textbf{7.51} &	\textbf{10.3} &	7.98 &	\textbf{11.72} & - \\
& Rnorm & 1.02 & 1.08 & 1.04 & \textbf{0.96} & \textbf{0.85} & 1.05 & \textbf{0.96} & \textbf{0.99}\\
\multirow{2}{*}
& AMF-V2022 \cite{liang2024amf} & 8.16	& 11.97 &  17.92 &	8.32 &	12.43 &	7.92 &	12.5 & - \\
& Rnorm  & 1.03 & 1.05 & 1.07 & 1.07 & 1.03 & 1.04 & 1.01 & 1.04 \\
\multirow{2}{*}
& TD-Placer-V2022 & \textbf{7.92} & \textbf{11.42} & \textbf{16.68} &	7.81 &	12.1 &	\textbf{7.59} &	12.41 & - \\
 & Rnorm & \textbf{1} & \textbf{1} & \textbf{1} & 1 & 1 & \textbf{1} & 1 & 1 \\
 
\addlinespace
\cline{2-10}
\addlinespace

\multirow{2}{*}
& V2023 \cite{Xilinx2023}& 7.86 & \textbf{11.29} & 17.78 &	\textbf{7.77} &	\textbf{10.91} &	8.26 &	\textbf{11.76}  & - \\
& Rnorm & 1.0 & \textbf{0.96} & 1.05 & \textbf{0.97} & \textbf{0.91} & 1.06 & \textbf{0.95} & \textbf{0.99} \\
\multirow{2}{*}
& AMF-V2023 \cite{liang2024amf} & 8.14 & 12.14 &	18.24 &	8.34 &	12.38 &	8.17 &	12.43 & - \\
& Rnorm & 1.04 & 1.04 & 1.08 & 1.05 & 1.03 & 1.05 & 1.01 & 1.04 \\
\multirow{2}{*}
& TD-Placer-V2023 & \textbf{7.83} &	11.71 &	\textbf{16.91} &	7.97 &	12.04 &	\textbf{7.78} &	12.34 & - \\
& Rnorm & \textbf{1} & 1 & \textbf{1} & 1 & 1 & \textbf{1} & 1 & 1 \\

\addlinespace
\cline{2-10}
\addlinespace

\multirow{2}{*}
& V2024  \cite{Xilinx2024}& \textbf{7.71}	 & \textbf{10.72} & 18.13 & \textbf{7.6}	 & \textbf{10.26} & \textbf{7.56}	 & \textbf{12.04} & - \\
& Rnorm & \textbf{0.97} & \textbf{0.94} & 1.05 & \textbf{0.91} & \textbf{0.88} & \textbf{1.0} &  \textbf{0.98} & \textbf{0.96} \\
\multirow{2}{*}
& AMF-V2024 \cite{liang2024amf} & 8.21	 & 11.8	 & 17.96 & 8.34	 & 12.24 & 8.2 &	12.47 & - \\
& Rnorm & 1.04 & 1.04 & 1.05 & 1.04 & 1.05 & 1.08 & 1.01 & 1.04 \\
\multirow{2}{*}
& TD-Placer-V2024 & 7.91 &	11.35 &	\textbf{17.2} &	8.02 &	11.7 &	\textbf7.58 &	12.32 & - \\
& Rnorm & 1 & 1 & \textbf{1} & 1 & 1 & 1 & 1 & 1 \\

    \bottomrule
    \end{tabular}%

  \end{threeparttable}
  \label{expriment1.1}%
\end{table*}

\begin{table}[!t]
\setlength{\abovecaptionskip}{0pt}
\setlength{\belowcaptionskip}{0pt}
\caption{ \textsc{ Comparison of Regression Models }}
\begin{center}
\begin{threeparttable}
\renewcommand{\arraystretch}{1.5}  
\begin{tabular}{c@{\hskip 15pt}c@{\hskip 15pt}c@{\hskip 15pt}c}
\toprule
\textbf{Regression Model} & \textbf{MAE} & \textbf{RMSE} & \textbf{R²} \\
\midrule
GBR         & 0.198 & 0.314 & 0.6  \\
RF          & 0.178 & 0.279 & 0.63 \\
AMF-Placer \cite{liang2024amf} & 0.119 & 0.186 & 0.71 \\
\textbf{TD-Placer-DP}        & \textbf{0.097} & \textbf{0.171} & \textbf{0.8}  \\
\bottomrule 
\end{tabular}
\end{threeparttable}
\label{expriment2}
\end{center}
\end{table}

\subsection{\textbf{The accuracy of the end-to-end delay prediction}}


\textbf{Motivation:} We evaluate TD-Placer’s end-to-end delay prediction (referred to as the TD-Placer-DP) accuracy on the dataset from Section \ref{datasettvt}, comparing it with Random Forest (RF), Gradient Boosting Regression (GBR), and AMF-Placer’s piecewise polynomial model\cite{liang2024amf}. Hyperparameters are tuned via grid search on the validation set for optimal comparison.

\textbf{Results: }The performance comparison of different regression models on the same test set is presented in \ref{expriment2}. Specifically, TD-Placer-DP achieves the best generalization performance, with a Mean Absolute Error (MAE), Root Mean Square Error (RMSE), and R² of 0.097, 0.171, and 0.80, respectively. The piecewise linear regression model adopted in AMF-placer ranks second, with MAE, RMSE, and R² values of 0.119, 0.186, and 0.71. In contrast, Random Forest and Gradient Boosting Regression show inferior generalization ability, with their respective metrics being 0.178, 0.279, 0.63 and 0.198, 0.314, 0.60. The limited generalization of models like RF and GBR arises from their reliance on single-feature splits, which struggle to capture complex feature interactions.
In contrast, TD-Placer-DP provides a more hierarchical and interpretable approach to feature modeling, enabling it to capture complex relationships among multiple timing-related features. As a result, It achieves the best performance, surpassing the second-best AMF-Placer model by 23\%, 9\%, and 13\% across the three metrics.

\begin{table}[!t]
\setlength{\abovecaptionskip}{0pt}
\setlength{\belowcaptionskip}{0pt}
\caption{ \textsc{Ablation of Timing-Related Features} }
\begin{center}
\begin{threeparttable}
\renewcommand{\arraystretch}{1.5}  
\begin{tabular}{c@{\hskip 15pt}c@{\hskip 15pt}c@{\hskip 15pt}c}
\toprule
\textbf{Regression Model} & \textbf{MAE} & \textbf{RMSE} & \textbf{R²} \\
\midrule
Net vertex \& Net environment (W/O)         & 0.118 & 0.187 & 0.69  \\
Net environment (W/O)   & 0.107 & 0.18 & 0.72 \\
Net vertex (W/O)   & 0.105 & 0.178 & 0.74 \\
TD-Placer-DP (W)  & \textbf{0.097} & \textbf{0.171} & \textbf{0.8}  \\
\bottomrule 
\end{tabular}
\end{threeparttable}
\caption*{\small \textit{(W/O) indicates that the corresponding category of TD-Placer-DP features is removed; (W) indicates the full model using all feature categories.}}
\label{expriment3}
\end{center}
\end{table}

\subsection{\textbf{Timing-related features ablation}}

\textbf{Motivation:} In this section, we investigate the contribution of three types of features to net delay prediction: (a) net vertex features, (b) net environment features, and (c) pin routing features. We remove each feature type individually while keeping the model size comparable, and ensure that all models are trained and tested on the same training and test sets to ensure a fair comparison.

\textbf{Results:} The effectiveness of different combinations of net vertex features, net environment features, and pin routing features for delay prediction is summarized in Table \ref{expriment3}. Specifically, when net vertex features and net environment features are removed, TD-Placer-DP achieves MAE, RMSE, and R² scores of 0.118, 0.187, and 0.69 on the test set.
After adding net vertex features, the performance improves to 0.107, 0.180, and 0.72, respectively, indicating that considering net load and crosstalk coupling effects significantly benefits delay prediction. When net vertex features are removed and only pin routing and net environment features are used, the metrics further improve to 0.105, 0.178, and 0.74, suggesting that net environment also plays a crucial role in delay prediction. Finally, when all three categories of features are used together, TD-Placer-DP achieves the best generalization performance, with MAE, RMSE, and R² reaching 0.097, 0.171, and 0.80, respectively. 
This observation implies that both the net’s structural topology and its surrounding environment contribute complementary and delay-relevant information for capturing the timing behavior between the driver and load pins. 

\subsection{\textbf{Timing model evaluation}}

\begin{table*}[!t]
  \centering
  \caption{\textsc{ Timing Model Validation With Benchmarks }}
  \begin{threeparttable}
    \renewcommand{\arraystretch}{1.1} 
    \begin{tabular}{
      >{\centering\arraybackslash}m{4.6cm}  
      >{\centering\arraybackslash}m{1.0cm}  
      >{\centering\arraybackslash}m{1.0cm}  
      >{\centering\arraybackslash}m{1.0cm}  
      >{\centering\arraybackslash}m{1.0cm}  
      >{\centering\arraybackslash}m{1.0cm}  
      >{\centering\arraybackslash}m{1.0cm}  
      >{\centering\arraybackslash}m{1.0cm}  
      >{\centering\arraybackslash}m{1.0cm}  
    }
    \toprule
    Benchmark  & BLSTM & DigitRecog & FaceDetect  & SpooNN & MemN2N & MiniMap2 & OpenPiton & Average \\
\midrule
\multirow{2}{*}{}
Vivado Pre-Route CPD Prediction \cite{Xilinx2024} & \textbf{8.57} & \textbf{11.68} & \textbf{17.87} & \textbf{8.24} & \textbf{11.76} & 8.11 &
12.87 & -\\
Relative Error(\%) & 8.3\%	& 3.0\%	& 3.9\%	&3.0\% &1.0\% & 7.0\%	&4.4\%	 &\textbf{4.4}\%	
 \\
\midrule
\multirow{2}{*}{}
AMF CPD Prediction \cite{liang2024amf}& 6.93 & 9.59 & 14.38 & 9.51 & 10.77 & 7.38 &
11.55 & -\\
Relative Error(\%) & 12.4\%	& 15.5\%	& 16.4\%	& 18.8\%	& 8.0\% & 2.7\%	& 6.3\%	& 11.4\% 
 \\
 \midrule
\multirow{2}{*}{}
TD-Placer-TM & 7.12 & 9.88 & 15.62 & 9.31 & 10.93 & \textbf{7.45} &
\textbf{11.82} & -\\
Relative Error(\%) & 10.0\%	& 13.0\%	& 9.2\%	& 13.8\%	& 6.6\% & 1.7\%	& 4.1\%	& 8.3\% 
 \\
 \midrule
Actual Post-Route CPD(ns) & 7.91	&11.35	&17.2	&8.02	&11.7 &7.58	&12.32	& -
 \\
\bottomrule
\end{tabular}%

  \end{threeparttable}
  \label{expriment4}%
\end{table*}

\begin{table*}[!t]
  \centering
  \caption{\textsc{Fine-gained weighting scheme ablation}}
  \begin{threeparttable}
    \renewcommand{\arraystretch}{1.0} 
    \begin{tabular}{
      >{\centering\arraybackslash}m{0.8cm}  
      >{\centering\arraybackslash}m{3.2cm}  
      >{\centering\arraybackslash}m{1.0cm}  
      >{\centering\arraybackslash}m{1.0cm}  
      >{\centering\arraybackslash}m{1.0cm}  
      >{\centering\arraybackslash}m{1.0cm}  
      >{\centering\arraybackslash}m{1.0cm}  
      >{\centering\arraybackslash}m{0.9cm}  
      >{\centering\arraybackslash}m{0.9cm}  
      >{\centering\arraybackslash}m{0.8cm}  
    }
    \toprule
    EVA & Method  & BLSTM & DigitRecog & FaceDetect  & SpooNN & MemN2N & MiniMap2 & OpenPiton & Average \\
    \midrule
\addlinespace
\multirow{4}{*}{CPD}
& TD-Placer (without dgw) & 8.03 & 11.6 & 17.63 & 8.67 & 11.81 & 7.76 & 12.62 & -\\
& Rnorm & 1.015  & 1.022	& 1.025	&1.081 &1.009 & 1.024	&1.024	 & 1.028 \\
\addlinespace
\cline{2-10}
\addlinespace
\multirow{2}{*}
& TD-Placer (with dgw) & \textbf{7.91} & \textbf{11.35} & \textbf{17.2} & \textbf{8.02} & \textbf{11.7} & \textbf{7.58} &
\textbf{12.32} & -\\
& Rnorm  & \textbf{1}	& \textbf{1}	& \textbf{1}	& \textbf{1}	&\textbf{1} & \textbf{1}	& \textbf{1}	& \textbf{1} \\
\addlinespace
\midrule
\addlinespace

\multirow{2}{*}{TNS}
& TD-Placer (without dgw) & -14 & -62242 & 0 & -23 & -1563 & 0 & -23247 & -\\
\addlinespace
\cline{2-10}
\addlinespace
& TD-Placer (with dgw) & \textbf{-8} & \textbf{-61307} & \textbf{0} & \textbf{-12} & \textbf{-1522} & \textbf{0} &
\textbf{-21363} & -\\
    \bottomrule
    \end{tabular}%

  \end{threeparttable}
  \label{expriment5.0}%
\end{table*}

\textbf{Motivation: }In this section, we evaluate the accuracy of the timing analysis model embedded in TD-Placer (referred to as the TD-Placer-TM) in predicting the actual post-routing critical path delay.  We apply TD-Placer-TM, the  the timing model used in AMF-Placer, and Vivado 2024.2 to predict critical path delays on the placed benchmarks, and compare them with actual post-routing delays reported by Vivado.

\textbf{Results: }Table \ref{expriment4} shows the results. TD-Placer-TM slightly lags behind Vivado 2024.2 in post-routing critical path delay prediction but outperforms AMF-Placer. The average error across seven benchmarks is 8.3\% for TD-Placer-TM, 4.4\% for Vivado, and 11.4\% for AMF-Placer. Vivado benefits from internal knowledge of its routing and device-specific timing. TD-Placer-TM’s estimates are optimistic due to two main factors:
\begin{itemize}
\item \textbf{The impact of routing congestion:} Most placed instances lie in uncongested regions, resulting in a congestion-imbalanced training set. Although TD-Placer-TM is congestion-aware, the dominance of uncongested samples biases learning toward the majority, leading to delay underestimation in congested areas where critical paths are more likely to pass.
\item \textbf{The variability in logic delay estimation: }Logic delay is estimated via a lookup table. While per-instance error is small, it accumulates along the critical path. For example, in the BLSTM benchmark, 14 instances with 0.015 ns error each lead to a 0.2 ns total, accounting for a large portion of the 0.8 ns gap between predicted and post-routing delay.
\end{itemize}

\subsection{\textbf{Fine-gained weighting scheme ablation}}

\textbf{Motivation:} This section evaluates the effectiveness of the depth- and global-timing-aware weighting scheme (referred
to dgw). We perform placement with and without this scheme in TD-Placer, then conduct routing using Vivado 2024.2. The post-routing timing quality of the designs is compared.

\textbf{Result:} As shown in Table \ref{expriment5.0}, removing dgw scheme leads to an average post-routing CPD increase of 2.8\% across seven benchmarks. Moreover, TNS exhibits a significant degradation. These results indicate that dgw effectively guides optimization of critical timing paths during global placement iterations without imposing additional overhead on low-risk paths, thereby enhancing the benefits of accurate timing model.

\section{\textbf{Conclusion and Future Work}}
This paper presents a timing-driven global placement framework, TD-Placer, that leverages global net information and timing features to predict delays and apply fine-grained weights for smooth critical path delay reduction. Compared to state-of-the-art placers, it improves WNS and CPD by $ \sim $10\% and $ \sim $5\%, respectively, with CPD closely matching the average of five Vivado versions (2020.2–2024.2) within 1\% ($\times1.01$).

Future work may integrate clock-skew-aware placement and finer logic-delay handling to further optimize timing. Overall, TD-Placer provides a scalable deep learning paradigm for timing-driven placement and a promising direction for future research.

\bibliographystyle{IEEEtran}
\bibliography{1-TCAD/placementopgy/Placement_Opt_Ref}

\end{document}